\documentclass[twocolumn,twocolappendix]{aastex701}
\usepackage{xspace}

\newcommand\mesa{\textsf{MESA}\xspace}
\newcommand\gyret{\textsf{GYRE-tides}\xspace}
\newcommand\gyre{\textsf{GYRE}\xspace}

\newcommand{\IM}{\mathrm{Im}}
\defcitealias{hut_tidal_1981}{H81}
\defcitealias{hurley_evolution_2002}{HTP02}

\usepackage{CJKutf8}
\usepackage{amsmath}
\usepackage{hyperref}
\usepackage{tikz}
\usetikzlibrary{arrows.meta,decorations.markings,patterns,calc,positioning,fit,matrix}

\begin{document}

\shorttitle{Interior Radiative Damping Dominates the Tidal Evolution of TOI-5882}
\shortauthors{Narayan, Soares-Furtado, \& Townsend}

\title{The Tale of a Hungry Subgiant and Its Brown Dwarf: \\ Interior Radiative Damping Dominates the Tidal Evolution of TOI-5882}

\begin{CJK*}{UTF8}{gbsn}

\author[orcid=0009-0007-0488-5685]{Ritvik Sai Narayan}
\affiliation{Department of Astronomy, University of Wisconsin--Madison, 475 N.~Charter St., Madison, WI 53706, USA}
\affiliation{Wisconsin Center for Origins Research, University of Wisconsin--Madison, 475 N Charter St, Madison, WI 53706, USA}
\email[show]{ritviksn@mit.edu}  

\author[orcid=0000-0001-7493-7419,sname='Soares-Furtado']{Melinda Soares-Furtado}
\affiliation{Department of Astronomy, University of Wisconsin--Madison, 475 N.~Charter St., Madison, WI 53706, USA}
\affiliation{Wisconsin Center for Origins Research, University of Wisconsin--Madison, 475 N Charter St, Madison, WI 53706, USA}
\affiliation{Department of Physics, University of Wisconsin--Madison, 1150 University Avenue, Madison, WI 53706, USA}
\email{mmsoares@wisc.edu}  

\author[orcid=0000-0002-2522-8605]{Richard H. D. Townsend}
\affiliation{Department of Astronomy, University of Wisconsin--Madison, 475 N.~Charter St., Madison, WI 53706, USA}
\affiliation{Wisconsin Center for Origins Research, University of Wisconsin--Madison, 475 N Charter St, Madison, WI 53706, USA}
\email{townsend@astro.wisc.edu}

\begin{abstract}
We present a self-consistent tidal evolution framework that couples binary evolution from \textsf{MESA} to the full linear tidal response from \textsf{GYRE-tides}. Applying this framework to TOI-5882, a subgiant hosting a short-period brown dwarf, we show that interior radiative damping dominates the system's tidal evolution, with the classical equilibrium tidal model significantly underestimating the star's angular momentum evolution by several orders of magnitude. Consequently, our combined framework predicts a 2--6 fold reduction in the engulfment timescale, accelerating the companion's inspiral by roughly 25--110 Myr. By modeling angular momentum transport through the star as it evolves, we demonstrate that the early inspiral is driven by the non-resonant dissipation of internal gravity waves, before transitioning into a regime dominated by resonance crossings as the system approaches Roche-lobe overflow. We highlight the necessity of reframing the historical dichotomy between equilibrium and dynamical tides and instead propose categorizing tidal interactions around their dissipation mechanisms: radiatively damped tides, driven by radiative diffusion, and viscously damped tides, driven by turbulent viscosity. Our framework is broadly applicable to the tidal modeling of a wide class of star-companion systems, from binary stars to hot Jupiters, in a self-consistent and computationally feasible manner. 

\end{abstract}

\keywords{\uat{Tidal interaction}{1699} --- \uat{Stellar oscillations}{1617} --- \uat{Stellar interiors}{1606} --- \uat{Stellar rotation}{1629} --- \uat{Subgiant stars}{1646} --- \uat{Brown dwarfs}{185}}



\section{Introduction} \label{section:introduction}

The post-main-sequence evolution of stars with companions fundamentally reshapes their orbital architectures. 
As a star expands during the subgiant and red-giant phases, tidal effects and stellar mass loss drive dynamical pathways that can destabilize close-in companions, alter long-term orbital configuration, and ultimately trigger engulfment and common-envelope evolution events \citep[e.g.,][]{zahn_tidal_1977,verbunt_tidal_1995,penev_direct_2009,villaver_orbital_2009,veras_post-main-sequence_2016}. 
Understanding how and when such events occur is critical in modeling the formation of mass transfer products such as blue straggler stars \citep{mathieu_blue_2025} and the fate of planetary systems \citep{bonsor_post-main_2011}. As habitable zones migrate with the evolution of the host star, the post-main-sequence phase creates new regions of temperate irradiation while simultaneously engulfing interior planets through tidal decay \citep{danchi_evolution_2006,danchi_effect_2013,ramirez_habitable_2016}. Hence, the interplay between these processes determines which planets survive, migrate, or are engulfed \citep{rasio_tidal_1996}. 

Engulfment events provide a uniquely informative window into the coupled evolution of stars and planets. The engulfment of a closely orbiting companion can alter stellar spin \citep[e.g.,][]{oetjens_influence_2020,tayar_spinning_2022}, modify its surface chemistry \citep[e.g.,][]{soares-furtado_lithium_2021,behmard_planet_2023}, and deposit orbital energy into its envelope \citep[e.g.,][]{yarza_hydrodynamics_2023,oconnor_giant_2023}, thereby influencing subsequent stellar evolution. Hence, these effects make engulfment one of the few observable markers of star-companion interactions over Gyr timescales. 

Compact subgiant systems are especially valuable laboratories for studying these processes because they are at the threshold where tidal interactions can intensify and stellar evolution is rapid. TOI-5882, a $1.3 \, M_\odot$ subgiant hosting a $22\,M_{\rm Jup}$ brown dwarf (BD) at a 7.14-day period, represents a system that is on the brink of such dynamical processes \citep{vowell_eleven_2025}. Modeling the evolution of this system provides an opportunity to examine the effect of tidal decay, stellar evolution, and energy deposition across different timescales. 

Moreover, the eventual engulfment of TOI-5882b and the common envelope phase that follows could help inform the formation of short-period post-common-envelope systems. Prominent examples include WD 0137-349 \citep[$P_{\rm orb} \sim 2$\,hr;][]{burleigh_near-infrared_2006} and NLTT 5306B \citep[$P_{\rm orb} \lessapprox 72$\,min;][]{steele_nltt_2013}, which have BDs at periods of a few hours, and WD 1856+534b \citep[$P_{\rm orb} \sim 1.4$\,d;][]{vanderburg_giant_2020}, the only white dwarf currently known to host an intact planet. Determining whether TOI-5882b will survive to become a similar compact remnant requires a precise accounting of the forces driving its current inspiral. 
Due to the almost circular orbit of TOI-5882b ($e = 0.0339 \pm 0.0041$), this dynamic evolution is overwhelmingly driven by the tides raised on the primary star.

To accurately capture this regime and track the inspiral until Roche-lobe overflow (RLOF) of the primary, we develop a coupled \mesa--\gyret framework. In this approach, \mesa \citep[\textit{Modules for Experiments in Stellar Astrophysics};][]{paxton_modules_2011,paxton_modules_2013,paxton_modules_2015,paxton_modules_2018,paxton_modules_2019,jermyn_modules_2023} provides the 1-D treatment of stellar evolution and processes such as convection, rotation, mass loss, and binary interactions across evolutionary phases. At each evolutionary timestep, we pass these stellar models into \gyre \citep{townsend_gyre_2013,townsend_angular_2018,goldstein_contour_2020,sun_gyre_tides_2023}, an oscillation code that solves the linearized equations of nonadiabatic stellar pulsations. This real-time integration allows us to self-consistently compute the full tidal response and energy deposition within the primary star as it evolves.

This Letter is organized as follows: In Section~\ref{section:tides}, we introduce our tidal evolution model, detailing the underlying theoretical formalism and our new software implementation. In Section~\ref{section:stellar_model}, we establish the single-star and binary evolution setups used for TOI-5882. We present our results in Section~\ref{section:results}, and in Section~\ref{section:discussion}, we contextualize these findings by examining the limitations of the classical tide formalism and discussing broader applications of our framework. 

\section{Tidal Evolution Model} \label{section:tides}

\subsection{Motivation} \label{subsection:motivation}

Tidal interactions govern the exchange between rotational and orbital angular momentum in celestial systems. Classically, tidal theory has distinguished between the equilibrium tide, which describes the quasi-static stellar deformation in which energy is dissipated through frictional processes (i.e., the weak friction approximation) due to the presence of a tidal lag, and the dynamical tide, which results from the radiative damping of tidally-excited internal gravity waves (IGWs) \citep{zahn_dynamical_1975,zahn_tidal_1977,zahn_tidal_1989}. Given that the \citet{hut_tidal_1981} and \citet{hurley_evolution_2002} equilibrium tidal prescriptions (hereafter, \citetalias{hut_tidal_1981} and \citetalias{hurley_evolution_2002} respectively) describe a secular response that can be parameterized through a set of analytically solvable differential equations, they have been widely adopted in studies of orbital evolution for stars with convective envelopes\footnote{Recent works in the assessment of tidal migration of planets around white dwarfs also use the equilibrium tide formalism with a constant time lag model \citep[e.g.,][]{trani_ominous_2020,li_can_2025}.}. 
However, this approach neglects any contribution to tidal dissipation from deeper regions and the frequency-dependent coupling between the orbit and the star's internal mode spectrum. 
Accounting for these effects during the post-main-sequence evolution of stars could lead to the resonant enhancement of tidal dissipation \citep{ogilvie_tidal_2007} and the non-uniform deposition of angular momentum through the star \citep{fuller_angular_2014}. 

These effects can become important to accurately model the post-main-sequence evolution of stars, as their stellar structures can drastically change during these phases. Despite their recognized importance, tidal responses throughout the entire star are yet to be modeled alongside stellar evolution. Most existing frameworks adopt prescriptions that consider tidal dissipation only in stellar envelopes and are parameterized independently of evolving stellar structure \citep{villaver_orbital_2009,nordhaus_tides_2010}. An accurate assessment of tidal decay in post-main-sequence systems such as TOI-5882b therefore requires models that allow tidal responses to evolve alongside stellar evolution. 

Motivated by these limitations, we develop a self-consistent tidal evolution framework that couples stellar evolution models from \mesa with the full tidal response calculated using \gyret \citep{sun_gyre_tides_2023}. In this effort, our primary goal is to develop an implementation that can continuously update the evolution of orbital angular momentum and energy using these calculations within the \mesa binary timestep. 

\subsection{Theoretical Formalism} \label{subsection:formalism}

In this work, we use the same theoretical formalism as \citet{townsend_angular_2018} and \citet{sun_gyre_tides_2023}. For completeness, we present the key equations used in evaluating the changes to orbital angular momentum and energy. We adopt a non-rotating reference frame with a primary of mass $M$ and photospheric radius $R$ at the origin and a companion of mass $qM$. In the below equations, we use the time coordinate $t$ and spherical position coordinates ($r$, $\theta$, $\phi$) to denote radial, polar, and azimuthal components. Tides are raised on the primary by the forces arising from the companion's tidal potential,

\begin{equation} \label{eqn:tidal_potential}
\begin{split}
        \Phi_{\rm T} (\mathbf{r}; \,t) = - \varepsilon_{\rm T} \frac{GM}{R} & \sum_{\ell, m, k}  \bar{c}_{\ell, m, k} \left( \frac{r}{R} \right)^{\ell} \\
        & \times Y^m_\ell (\theta, \phi) \exp\left(-ik \mathcal{M} \right),
\end{split}
\end{equation}
where
\begin{equation}
    \varepsilon_{\rm T} \equiv q \left( \frac{R}{a} \right)^3,
\end{equation}
is a dimensionless parameter that determines the strength of the tidal forcing, $\bar{c}_{\ell, m,k}$ is a tidal expansion coefficient (see Equation (A1) in \citealt{sun_gyre_tides_2023}), $G$ is the gravitational constant, $a$ is the semi-major axis, $Y^m_{\ell}$ is a spherical harmonic, and $\mathcal{M}$ is the mean anomaly of the orbit. The triple summation, here and subsequently, extends over harmonic degrees $\ell \geq 2$, azimuthal orders $-\ell \leq m \leq \ell$ and Fourier indices $-\infty \leq k \leq \infty$.

The response of the primary to forcing by this potential is determined by linearizing the hydrodynamical equations about the equilibrium unforced state, and then solving the resulting system of linear differential equations as a boundary-value problem. To calculate the resulting secular torque on the primary, we generalize Equation~(6) of \citet{townsend_angular_2018} as
\begin{equation} \label{eqn:townsend+}
    \begin{split}
        \frac{\partial \mathcal{T}}{\partial r}  = &- \frac{\partial}{\partial r} \left( r^2 \rho_0 \int_0^{2\pi} \int_0^\pi r \sin \theta \; v'_\phi v'_r \sin \theta \; d\theta \; d\phi \right) \\
        & - r^2 \int_0^{2\pi} \int_0^\pi \rho' \frac{\partial}{\partial\phi} \left( \Phi' + \Phi_{\rm T} \right) \sin \theta \; d\theta \; d\phi.
    \end{split}
\end{equation}
Here, $v'_r$, $v'_\phi$, $\rho'$ and $\Phi'$ are the Eulerian (fixed-position) perturbations to the radial velocity, azimuthal velocity, density, and self-gravitational potential, respectively. Physically, this equation decomposes the instantaneous, differential (per-unit-radius) torque $\partial \mathcal{T}/\partial r$ into contributions arising from the divergence of the Reynolds stress (the first term on the right-hand side) and forcing by the combination of self-gravity and the tidal potential (the second term). It reduces to Equation~(6) of \citet{townsend_angular_2018} when the tidal potential is neglected.

\citet{sun_gyre_tides_2023} provide expressions for the space and time dependence of the perturbations ($v'_r$, $v'_\phi$, etc.) in terms of the set of radial eigenfunctions $\tilde{\xi}_{r;\ell,m,k}(r)$, $\tilde{\xi}_{h;\ell,m,k}(r)$, $\tilde{\rho}'_{\ell,m,k}(r)$, $\tilde{\Phi}'_{\ell,m,k}(r)$ that are calculated by \gyre. Using these expressions, we average Equation~(\ref{eqn:townsend+}) over one orbital period to obtain the secular differential torque as
\begin{equation} \label{eqn:differential_torque}
\begin{split}
    \left\langle \frac{\partial \mathcal{T}}{\partial r} \right\rangle & \equiv 
    \frac{1}{2\pi} \int_{0}^{2\pi} \frac{\partial \mathcal{T}}{\partial r} \; d\mathcal{M} \\
    & 
    \begin{split}= \sum_{\ell, m, k} {\rm i}m \Bigg[ - \frac{d}{dr} \left(r^3 \rho_0 \, \sigma_{m,k}^2 \, \tilde{\xi}_{{\rm h}; \ell,m,k} \, \tilde{\xi}_{{\rm r}; \ell,m,k}^* \right) \\
    + r^2 \tilde{\rho}'_{\ell,m,k} \left( \tilde{\Phi}_{\ell,m,k}^{'*} + \tilde{\Phi}_{{\rm T};\,\ell,m,k}^{*} \right) \Bigg],
    \end{split}
\end{split}
\end{equation}
where
\begin{equation} \label{eqn:tidal_potential_radial}
        \tilde{\Phi}_{{\rm T};\ell,m,k}(r) \equiv - \varepsilon_{\rm T} \frac{GM}{R} \bar{c}_{\ell, m, k} \left( \frac{r}{R} \right)^{\ell}
\end{equation}
and
\begin{equation} \label{eqn:sigma}
    \sigma_{m,k} = k \Omega_{\rm orb} - m \Omega_{\rm rot}
\end{equation}
is the tidal forcing frequency in the frame co-rotating with the primary at angular velocity $\Omega_{\rm rot}$.
Integrating Equation (\ref{eqn:differential_torque}) up to the surface boundary radius $r_s$, we have the net torque on  the star in a form equivalent to Equation (23) of \citet{sun_gyre_tides_2023}, 
\begin{subequations} \label{eqn:torque}
\begin{equation} 
\begin{split} \label{subeqn:tidal_torque}
    \langle \mathcal{T} \rangle = 
    4q^2 \frac{GM^2}{a} \sum_{\ell, m, k \geq 0} \left(\frac{R}{a} \right)^{\ell+3} \left(\frac{r_s}{R} \right) ^{\ell + 1} \\ 
    \times \; \kappa_{\ell,m,k} \, \IM (\bar{F}_{\ell,m,k}) \bar{G}^{(4)}_{\ell,m,k},
\end{split}
\end{equation}
\begin{equation} \label{subeqn:gbar4}
    \bar{G}^{(4)}_{\ell,m,k} \equiv m \frac{2\ell + 1}{4\pi} \left( \frac{R}{a} \right)^{-\ell + 2} |\bar{c}_{\ell,m,k}|^2,
\end{equation}
\end{subequations}

\noindent where 
\begin{equation} \label{eqn:Fbar}
    \bar{F}_{\ell,m,k} \equiv  \frac{1}{2} \frac{\tilde{\Phi}'_{\ell,m,k} (r_s)}{\tilde{\Phi}_{{\rm T};\ell,m,k} (r_s)},
\end{equation}
is the normalized response function, and $\kappa_{\ell,m,k}$ is the mode-pairing symmetry factor (see Equation (53) of \citealt{willems_energy_2010}). Note that the summation over $k$ is now restricted to non-negative values.

To find the net energy deposition in the star, we can integrate the local work done by the tidal potential on a differential fluid element, over the entire stellar volume. As a shortcut, however, we can obtain the same result by using the torque-power relation from \citet{ogilvie_tidal_2014} that lets us replace $m$ with $k\Omega_{\rm orb}$ in Equation~(\ref{subeqn:gbar4}). This lets us reproduce Equations (5) and (6) from \citet{sun_numerical_2025}, 
\begin{subequations} \label{eqn:power}
\begin{equation} 
\begin{split} \label{subeqn:power_dissipated}
     \langle \mathcal{P} \rangle = 
    4q^2 \frac{GM^2}{a} \sum_{\ell, m, k \geq 0} \left(\frac{R}{a} \right)^{\ell+3} \left(\frac{r_s}{R} \right) ^{\ell + 1} \\ 
    \times \; \kappa_{\ell,m,k} \,\IM (\bar{F}_{\ell,m,k}) \bar{G}^{(5)}_{\ell,m,k},
\end{split}
\end{equation}
\begin{equation} \label{subeqn:gbar5}
    \bar{G}^{(5)}_{\ell,m,k} \equiv k \Omega_{\rm orb} \frac{2 \ell + 1}{4\pi} \left( \frac{R}{a} \right)^{-\ell +2} |\bar{c}_{\ell,m,k}|^2.
\end{equation}
\end{subequations}

\subsection{Software Implementation} \label{subsection:implementation}

\citet{paxton_modules_2015} introduced an in-memory framework to couple \gyre with \mesa, originally designed to facilitate single-star asteroseismology. By passing the instantaneous stellar structure to \gyre at each timestep, this framework allowed for the rapid computation of pulsation mode eigenfrequencies and eigenfunctions. Recently, \citet{sun_numerical_2025} validated the ability of \gyret to compute tidal responses with nonadiabatic pulsations, analyze combined radiative and convective damping, and compute wave luminosity at the radiative-convective boundary (RCB) in the case of WASP-12. In this work, we extend the current in-memory framework to use parameters that are unique to the binary module (e.g., $q$, $e$, and $\Omega_{\rm rot}$), thereby coupling the latest stable versions of \mesa (r26.04.1) and \gyre (9.0) to self-consistently model tidal evolution in stellar systems.

In this approach, the computation of the stellar tidal response and the secular evolution of the orbit are separated. \gyret is used to evaluate the linear tidal response of a star (or planet) to its companion, whereas the assembly of the tidal torque and energy deposition rates are handled within \mesa. 

The binary evolution proceeds through the standard \mesa timestep loop. At initialization, a subroutine registers the hook functions that allow externally computed contributions to be supplied to \mesa. In particular, the hooks {\tt other\_jdot\_ls} and {\tt other\_edot\_tidal} are used to provide the tidal contribution to the orbital angular momentum and energy, respectively. 

The evaluation of the linear tidal response can be performed at a user-defined cadence rather than at every timestep to provide computational flexibility. When this cadence condition is met, a snapshot of the stellar structure is exported along with the tidal forcing parameters to a wrapper module that constructs a corresponding \gyre model. The forced oscillation response is then solved for each $\{\ell, m, k\}$ and the outputs are used to calculate the quantities in Equations (\ref{eqn:torque}) and (\ref{eqn:power}). 

To stay consistent with \mesa convention and allow these quantities to represent changes to the orbit, the hook {\tt other\_jdot\_ls} is defined as, 
\begin{equation}
    \dot{J}_{\rm ls} = - \langle \mathcal{T} \rangle_{\rm primary} - \langle \mathcal{T} \rangle_{\rm secondary},
\end{equation}
and the hook {\tt other\_edot\_tidal} is defined as, 
\begin{subequations}
\begin{equation} 
\begin{split} \label{subeqn:edot}
    \dot{E}_{\rm tidal} = - \frac{1}{|E_{\rm orb}|} (\langle \mathcal P \rangle_{\rm primary} + \langle \mathcal P \rangle_{\rm secondary}),
\end{split}
\end{equation}
\begin{equation} \label{subeqn:Eorb}
   E_{\rm orb} = - q\frac{GM^2}{2a}.
\end{equation}
\end{subequations}

These quantities are cached and supplied to the \mesa binary solver until the next tidal response update is performed. The secondary components to both $\dot{J}_{\rm ls}$ and $\dot{E}_{\rm tidal}$ may be turned off in the case that the tides raised on the primary dominate the tidal evolution (as is the case for TOI-5882).

\section{(Sub)Stellar Evolution Model} \label{section:stellar_model}

\subsection{Single-Star Evolution} \label{subsection:single_star_evol}

\begin{figure*}[t!]
        \includegraphics[width=\linewidth]{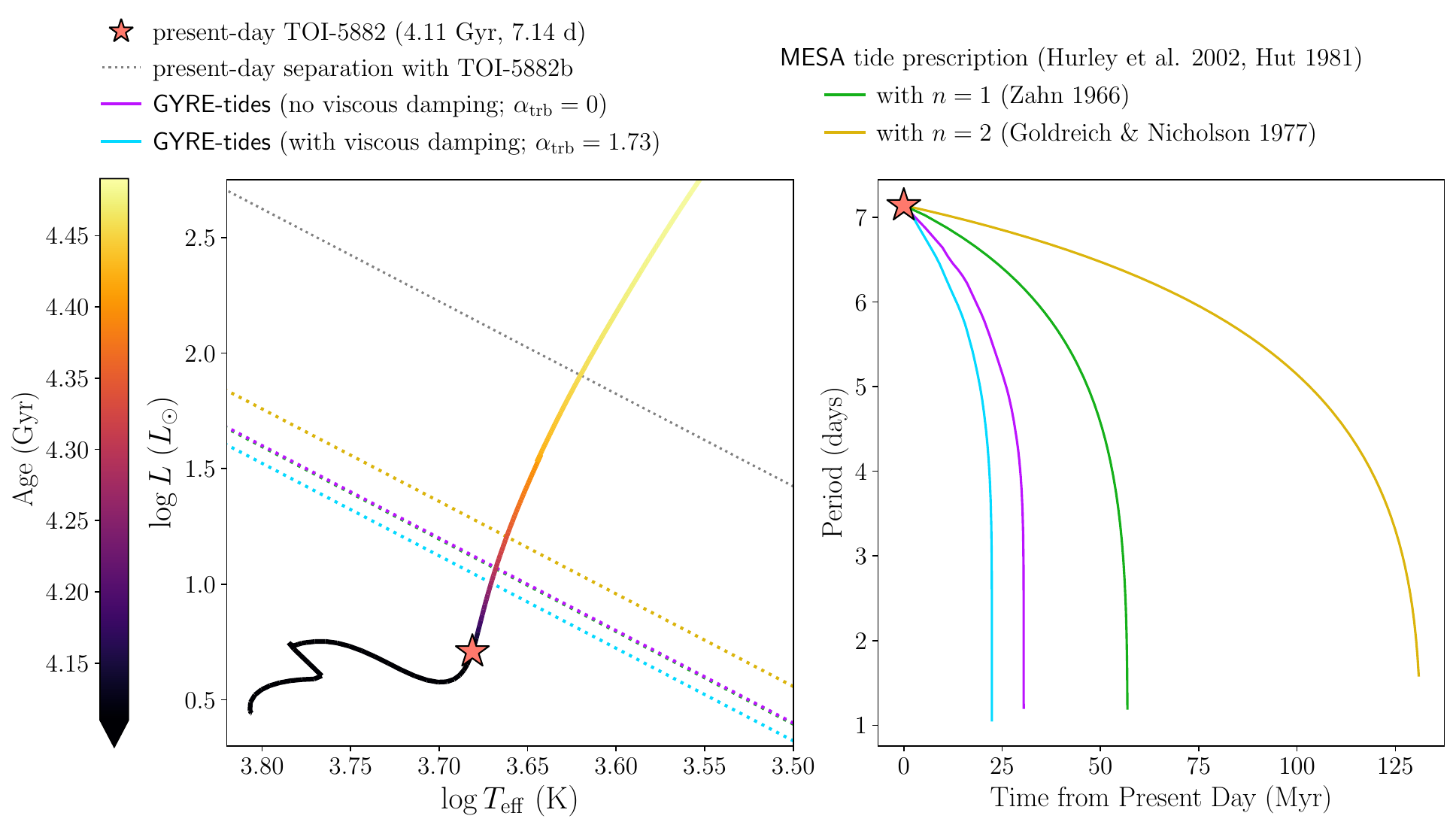}
         \caption{\textbf{Left:} Hertzsprung--Russell diagram showing the \mesa single-star evolutionary track of TOI-5882 colored by age (colorbar in Gyr), starting from the zero-age main sequence and evolving to the tip of the red giant branch. The present-day position of TOI-5882 is marked with a red star. Diagonal dotted lines of constant radii indicate the present-day orbital separation of TOI-5882b (gray) alongside the stellar radii at the onset of Roche-lobe overflow (RLOF) as predicted by the differing tidal prescriptions. \textbf{Right:} Orbital period decay as a function of time from the present day for the four different tidal prescriptions.}
        \label{fig:hr_diagram}
    \end{figure*}

We first establish a baseline using single-star evolutionary tracks from \mesa. While the \mesa binary framework evolves stars concurrently with the orbit, a single-star treatment can provide counterfactual epochs of evolution that would occur in the absence of a companion. This baseline is therefore useful in isolating deviations from single-star evolution in different formalisms. In the left panel of Figure~\ref{fig:hr_diagram}, we show the resulting Hertzsprung--Russell diagram evolved from the zero-age main sequence. 

We adopt a standard 1-D treatment of convection and mixing, with a mixing length parameter $\alpha_{\rm MLT} = 1.73$, which is empirically motivated by the asteroseismic modeling of subgiant stars \citep{hon_asteroseismic_2020}. We use a simple exponential convective overshoot prescription (with $f_0 = 0.004$ and $f = 0.014$) and apply a \citet{reimers_circumstellar_1975}-type wind ($\eta = 0.5$) during the red giant branch (RGB) evolution. We adopt these parameters from the \mesa test suite case \texttt{7MS\_prems\_to\_AGB} \citep{paxton_modules_2019}, substituting the target parameters appropriate for TOI-5882, and evolving the star to the tip of the red giant branch. 

We also include solid-body rotation and the star's observed metallicity \citep[$v \sin i = 7.3 \pm 0.5 \; {\rm km\; s^{-1}}$, $i = 88.56^{\circ}{}^{+0.97}_{-1.1}$, and $\textrm{[Fe/H] = 0.18}^{+0.16}_{-0.15}$;][]{vowell_eleven_2025} in our models. As a first endeavor into coupling detailed stellar evolution with a full tidal-response informed orbital evolution, we enforce uniform rotation throughout the star's evolution. While we do analyze the non-uniform radial profile of the tidal response and the localized deposition of angular momentum in post-processing (as detailed in Section~\ref{subsection:angular_momentum}), dynamically coupling this resulting differential rotational profile back into the evolutionary framework remains challenging. Thus, we account for rotation only via the shift of the wave frequencies in the corotating frame (Equation (\ref{eqn:sigma})), neglecting additional dynamical effects such as the Coriolis force and centrifugal distortion. To facilitate direct comparisons with interactions in the binary models, we keep the underlying single-star physics for the primary fixed across runs. All inlists required to reproduce our results with \mesa have been made publicly available via \textsf{Zenodo}\footnote{doi:10.5281/zenodo.18855026}. 

\subsection{Binary Evolution} \label{subsection:binary_evol}

We model the evolution of TOI-5882 using the \mesa binary module, with the primary being initialized from a saved stellar model, as described in Section~\ref{subsection:single_star_evol}, at the present-day system age of 4.11 Gyr. Similarly, we initialize the BD companion from a pre-computed substellar model of the same age (modified for TOI-5882b from the test-suite case \texttt{make\_brown\_dwarf}) using low-temperature opacity tables appropriate for substellar objects \citep{freedman_gaseous_2014}. The BD's rotation rate is assumed to be synchronized to the orbital frequency, given the close present-day separation. 

In the standard equilibrium tide prescription of \citetalias{hut_tidal_1981} and \citetalias{hurley_evolution_2002} used natively by \mesa, turbulent viscosity is suppressed when the half the tidal pumping timescale, $P_{\rm tid} = |\Omega_{\rm orb} - \Omega_{\rm rot}|^{-1}$, is shorter than the turnover timescale corresponding to the largest eddies, $\tau_{\rm conv}$ (see Equation (31) of \citetalias{hurley_evolution_2002}). This results in a global scaling factor for the turbulent viscosity, 
\begin{equation}
    f_{\rm conv} = \min \left[ 1, \left( \frac{|\Omega_{\rm orb} - \Omega_{\rm rot}|^{-1}}{2 \tau_{\rm conv}} \right)^n \right],
\end{equation}
where $n$ is the tidal reduction factor, which dictates the frequency-scaling of the viscosity reduction. 

In contrast, \gyret treats convective damping by adding a radial viscous force to the momentum equation \citep{willems_energy_2010,sun_numerical_2025}. This results in an effective turbulent viscosity scaling factor given by, 
\begin{equation} \label{eqn:gyretides_visc}
    f_{\rm GYRE} =  \left[ 1 + \left( \frac{\tau_{\rm ed} \sigma}{2\pi} \right)^n \right]^{-1},
\end{equation}
where the tidal reduction factor is set to one and $\tau_{\rm ed}$ is the local eddy-turnover timescale, determined by, 
\begin{equation}
    \tau_{\rm ed} = \frac{H\alpha_{\rm trb}}{v_{\rm ed}},
\end{equation}
with $H$ as the local pressure scale height, $v_{\rm ed}$ as the mixing-length theory convective velocity from our \mesa model, and $\alpha_{\rm trb}$ is a free parameter that is typically set equal to the mixing-length parameter $\alpha_{\rm MLT}$.

To isolate the effects of these differing treatments, we perform four binary evolution calculations: 
\begin{enumerate}
    \item The standard \mesa equilibrium tide prescription adopting a linear tidal reduction factor ($n = 1$), as described by \citet{zahn_marees_1966}. 
    \item The \mesa equilibrium tide prescription adopting a quadratic tidal reduction factor ($n=2$), as described by \citet{goldreich_turbulent_1977}. This prescription is the default \mesa treatment. 
    \item The \gyret framework with viscous damping disabled ($\alpha_{\rm trb} = 0$).
    \item The \gyret framework with both radiative and viscous damping, setting the turbulent viscosity efficiency equal to the mixing-length parameter ($\alpha_{\rm trb} = \alpha_{\rm MLT} = 1.73$).
\end{enumerate}

Given the mass ratio of the system and the circular orbit, we apply the \gyret calculation only for the primary ($\langle \mathcal T \rangle_{\rm secondary} = \langle \mathcal P \rangle_{\rm secondary} = 0$) and consider the quadrupole tidal harmonic ($\ell  = 2$; $|m| \leq 2$), with Fourier indices $-50 < k < 50$. While the contribution from $k \gtrapprox 2$ drops off rapidly for a circular orbit \citep{wu_hansen_2024}, we include the higher-order terms for completeness. Linear tidal theory requires that the fluid displacement amplitude remains small compared to its wavelength \citep[$k_r \tilde{\xi}_r \lesssim 1$;][]{goodman_dynamical_1998}. This assumption thus breaks down when the companion is sufficiently massive and the orbit is tight enough to initiate non-linear wave breaking \citep{barker_internal_2010}. For TOI-5882, we therefore evolve the system only until the primary fills its Roche lobe (computed using the fit of \citealt{eggleton_aproximations_1983}), and refer to the time from present day for the star to fill its Roche lobe as the RLOF timescale.

\section{Results} \label{section:results}

\subsection{Orbital Decay} \label{subsection:orbital_decay}

All four calculations described in Section~\ref{subsection:binary_evol} predict that the orbit decays monotonically as the primary ascends the RGB, with unstable mass transfer (given the unequal mass ratio) at the onset of RLOF. While \mesa's binary framework currently provides user hooks to update orbital energy and angular momentum, it lacks a direct interface to supply an externally computed eccentricity derivative ($\dot{e}$). Consequently, we cannot self-consistently evolve the orbital eccentricity using the full tidal response. 
However, because the present-day orbit of TOI-5882b is almost circular, we find that this current limitation has a negligible impact on our results.
More specifically, test integrations with and without \mesa's native circularization scheme enabled produce identical inspiral trajectories and RLOF timescales for each tidal prescription. 

The right panel of Figure~\ref{fig:hr_diagram} illustrates the divergence in the orbital evolution pathways predicted by the varying tidal prescriptions. While all models show an accelerating inspiral as the star evolves, the classical equilibrium tide formalism from \citetalias{hut_tidal_1981} and \citetalias{hurley_evolution_2002}, assuming dissipation only in the convective envelope, significantly underestimates the tidal dissipation rate in comparison to our framework. 

The $n=2$ tide prescription (gold solid line) yields the slowest orbital decay with the companion surviving for ${\sim}130$ Myr before RLOF. The $n=1$ prescription (green solid line) accelerates this to ${\sim}58$ Myr. Contrastingly, evaluating the full linear tidal response with \gyret predicts a much shorter timescale to RLOF, demonstrating that the system's evolution is heavily dominated by radiative damping. By setting $\alpha_{\rm trb} = 0$ to isolate the radiative contribution (magenta solid line), we find that this mechanism alone is sufficient to drive RLOF in just ${\sim}30$ Myr. When viscous damping ($\alpha_{\rm trb} = 1.73$) due to turbulent viscosity is also introduced (cyan solid line), the combined dissipation further reduces the RLOF timescale to ${\sim}22$ Myr. This indicates that while viscous damping in the convective regions contributes to tidal decay, the radiative damping of IGWs acts as the primary physical driver of the tidal inspiral, resulting in roughly a 2--6 fold reduction in the RLOF timescale in comparison to the \citetalias{hurley_evolution_2002} formalism. 

\subsection{IGW Excitation and Propagation} \label{subsection:g-modes}

   \begin{figure}[b!]
        \includegraphics[width=\linewidth]{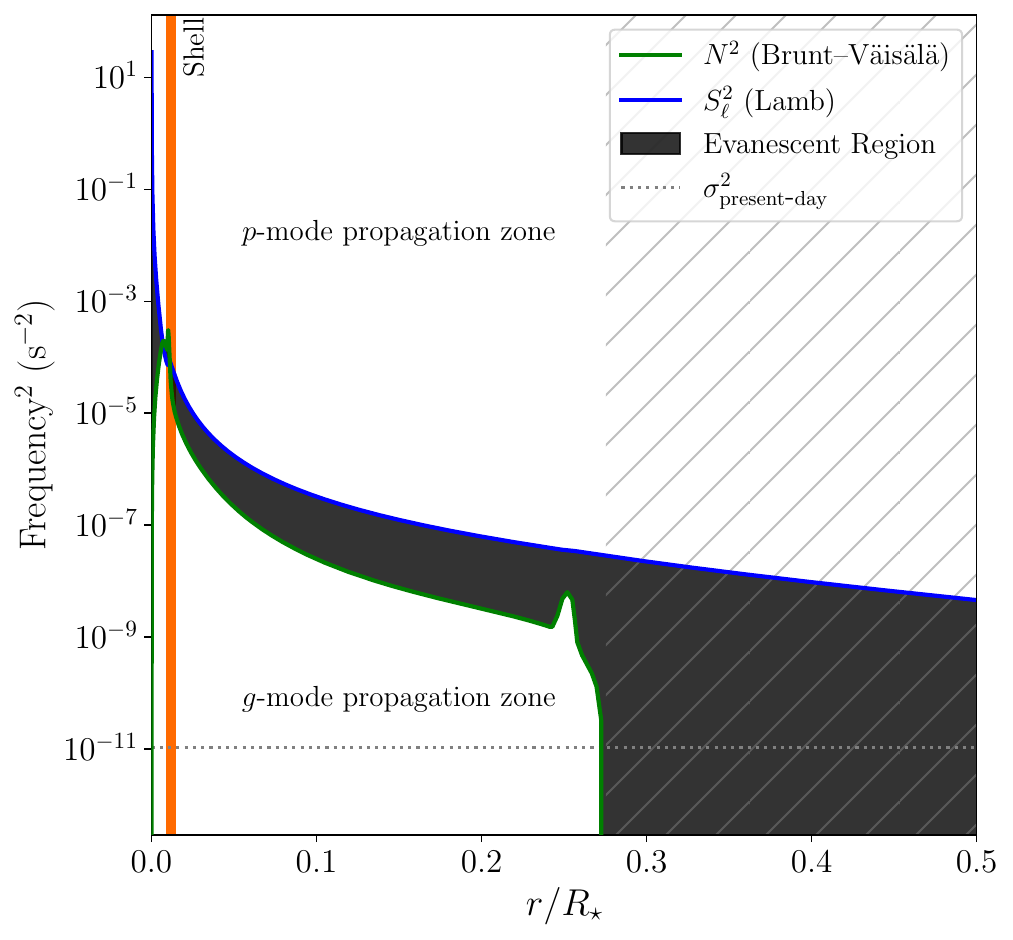}
        \caption{Propagation diagram for the primary TOI-5882 during its early RGB evolution. The green and blue solid curves show the radial profiles of the squared Brunt--Väisälä frequency and the $\ell=2$ Lamb frequency respectively. The dark shaded region denotes the evanescent region, which separates the outer acoustic ($p$-mode) cavity from the deep interior gravity ($g$-mode) cavity. The hatched region represents the convective region of the star while the orange shaded region denotes the radial location of the hydrogen burning shell. Lastly, the horizontal dotted line marks the present-day tidal forcing frequency induced by the BD companion. }
        \label{fig:propagation}
    \end{figure}

To further investigate the origin of the radiative-damping dominated inspiral, we examine the excitation and propagation of IGWs within the deeper regions of the star. The time-varying potential exerted by a massive companion can excite IGWs at the RCB \citep{zahn_dynamical_1975,goodman_dynamical_1998,terquem_tidal_1998,ogilvie_tidal_2007}. In Figure~\ref{fig:propagation}, we show a propagation diagram produced with a stellar model during the early RGB (also provided via Zenodo). The location of the horizontal dotted line shows that the present-day tidal forcing frequency lies well below both the Brunt--Väisälä frequency ($N$) and the quadrupole Lamb frequency ($S_2$), placing it comfortably within the interior $g$-mode propagation cavity. Consequently, IGWs are excited at the RCB and forced to propagate inward towards the core \citep{ivanov_unified_2013,bolmont_effect_2016,gallet_tidal_2017}. 

As these waves travel inward, they encounter a steeply rising Brunt--Väisälä frequency, which is sharply peaked at the hydrogen burning shell (denoted by the orange band in Figure~\ref{fig:propagation}). 
From Equation (3.368) in \citet{aerts_asteroseismology_2010}, the WKB radial wavenumber of IGWs (neglecting rotation) follows the relation:
\begin{equation} \label{eq:wavenumber}
    k_r^2 \sim \frac{N^2}{r^2\omega^2},
\end{equation}
where $\omega$ is the wave frequency. As $N$ peaks near the shell, the radial wavelength of the IGWs becomes small \citep{weinberg_nonlinear_2012, barker_tidal_2020,sun_gyre_tides_2023}. As demonstrated by Equation (19) of \citet{zahn_angular_1997}, this shortening can create steep temperature gradients between wave crests and troughs, which makes radiative damping extremely efficient in highly stratified regions. 

The strength of the IGWs and the extent to which radiative dissipation can damp them determine whether the waves can reflect in the core and form standing modes, instead of establishing the traveling wave regime \citep{goodman_dynamical_1998,barker_internal_2010}. If the tidally excited IGWs are in the standing wave regime, we would expect to find evidence for the star passing through tidal resonances during its evolution along the late SGB and RGB \citep{terquem_tidal_1998}. To determine which regime governs TOI-5882 during its current evolutionary phase, we must evaluate the system's tidal response as a function of the dimensionless forcing frequency $\sigma_{m,k}/\Omega_{\rm orb}$, where $\Omega_{\rm orb}$ is set to its present day value.  

    \begin{figure}[t!]
        \includegraphics[width=\linewidth]{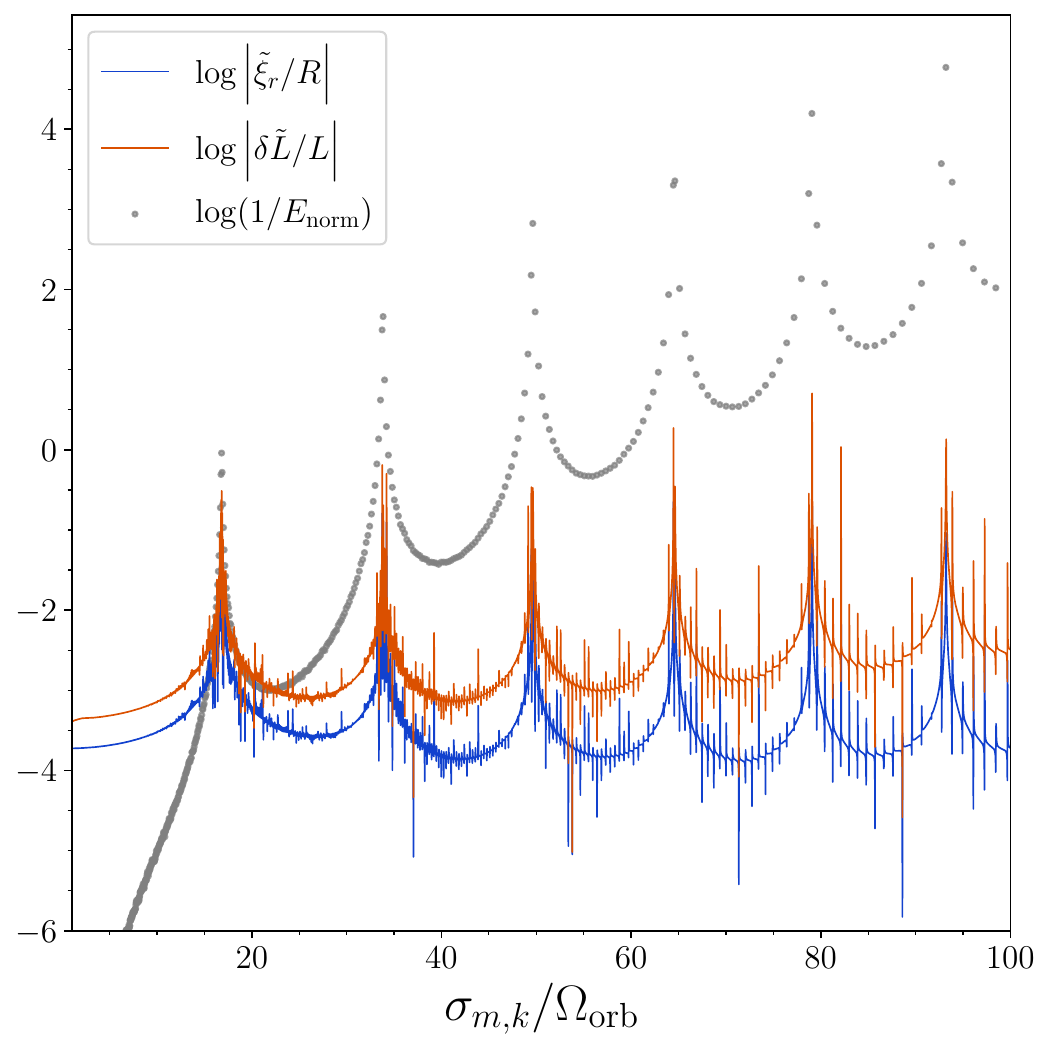}
        \caption{Tidal response of TOI-5882 as a function of the dimensionless forcing frequency $\sigma_{m,k}/\Omega_{\rm orb}$, computed with \gyret using $\alpha_{\rm frq}$ scanning continuously to sample the forcing frequency space. The blue and orange curves show the logarithmic amplitudes of the normalized radial displacement ($\log|\tilde{\xi}_r/R|$) and luminosity perturbation ($\log|\delta\tilde{L}/L|$), respectively, both evaluated at the stellar surface. Finally, gray points show the inverse normalized mode inertia from a free oscillation \gyre calculation spanning the same frequency range. We use the same stellar model as in Figure~\ref{fig:propagation} to perform these calculations. }
        \label{fig:resonances}
    \end{figure}

Since the tidal forcing frequency is only sampled at integer values of $k$, a direct \gyret calculation would undersample the true frequency space and miss narrow resonances. To rigorously search for any standing modes, we instead hold $k$, $m$, $\Omega_{\rm orb}$, and $\Omega_{\rm rot}$ constant and scan over the $\alpha_{\rm frq}$ parameter in the \texttt{\&tides} namelist. This acts as a scaling factor on the forcing frequency, allowing us to compute the surface amplitudes of radial displacement ($\tilde{\xi}_{\rm r}$) and luminosity perturbation ($\delta \tilde{L}$) at an arbitrarily high resolution. We present the tidal resonance spectrum in Figure~\ref{fig:resonances}.

To further probe this resonance structure, we perform a free oscillation \gyre calculation spanning the same frequency range. We overlay the star's inverse normalized mode inertia ($1/E_{\rm norm}$) in Figure~\ref{fig:resonances}, quantifying where in the star a given mode carries most of its energy \citep{dupret_theoretical_2009}. As demonstrated for red giant stars \citep{bedding_gravity_2011,mosser_probing_2012}, the coupling between the radiative interior and the convective envelope in TOI-5882 gives rise to mixed modes that carry signatures of $p$-modes and $g$-modes. As $p$-mode dominated mixed modes are confined to the low-density envelope, they possess low mode inertias (high $1/E_{\rm norm}$ peaks) and align precisely with the highest-amplitude resonances in the tidal response. 

For $e \simeq 0$, $|\bar{c}_{\ell,m,k}| = 0$ (see Section~\ref{subsection:formalism}) for $k > 2$ as the Hansen coefficients are mathematically forced to zero \citep{hughes_computation_1981}, with the quadrupole component carrying the largest amplitude. In the limit that $\Omega_{\rm rot} \ll \Omega_{\rm orb}$, the tidal response is concentrated to $\sigma/\Omega_{\rm orb} \simeq 2$. As seen in Figure~\ref{fig:resonances}, the tidal response near this frequency range is broad and relatively featureless rather than a sharp resonance peak. This behavior is consistent with the traveling wave regime, as the IGWs are highly damped before reaching the stellar center \citep{ogilvie_tidal_2007}. The early tidal evolution for TOI-5882 on the SGB and RGB thus arises from the non-resonant dissipation of IGWs in the stellar interior. However, as $\Omega_{\rm orb}$ increases, the resonance structure in terms of the dimensionless forcing frequency in Figure~\ref{fig:resonances} moves leftward, resulting in resonances near the onset of RLOF (see upper panel of Figure~\ref{fig:torque} for $\Omega_{\rm orb}/2\pi \gtrapprox 0.55$). To resolve the narrow widths of these resonance crossings in orbital frequency space, we restrict the \mesa timestep to 5\% of its default value.

\subsection{Angular Momentum Transport and Deposition} \label{subsection:angular_momentum}

\begin{figure*}[t!]
    \includegraphics[width=\linewidth]{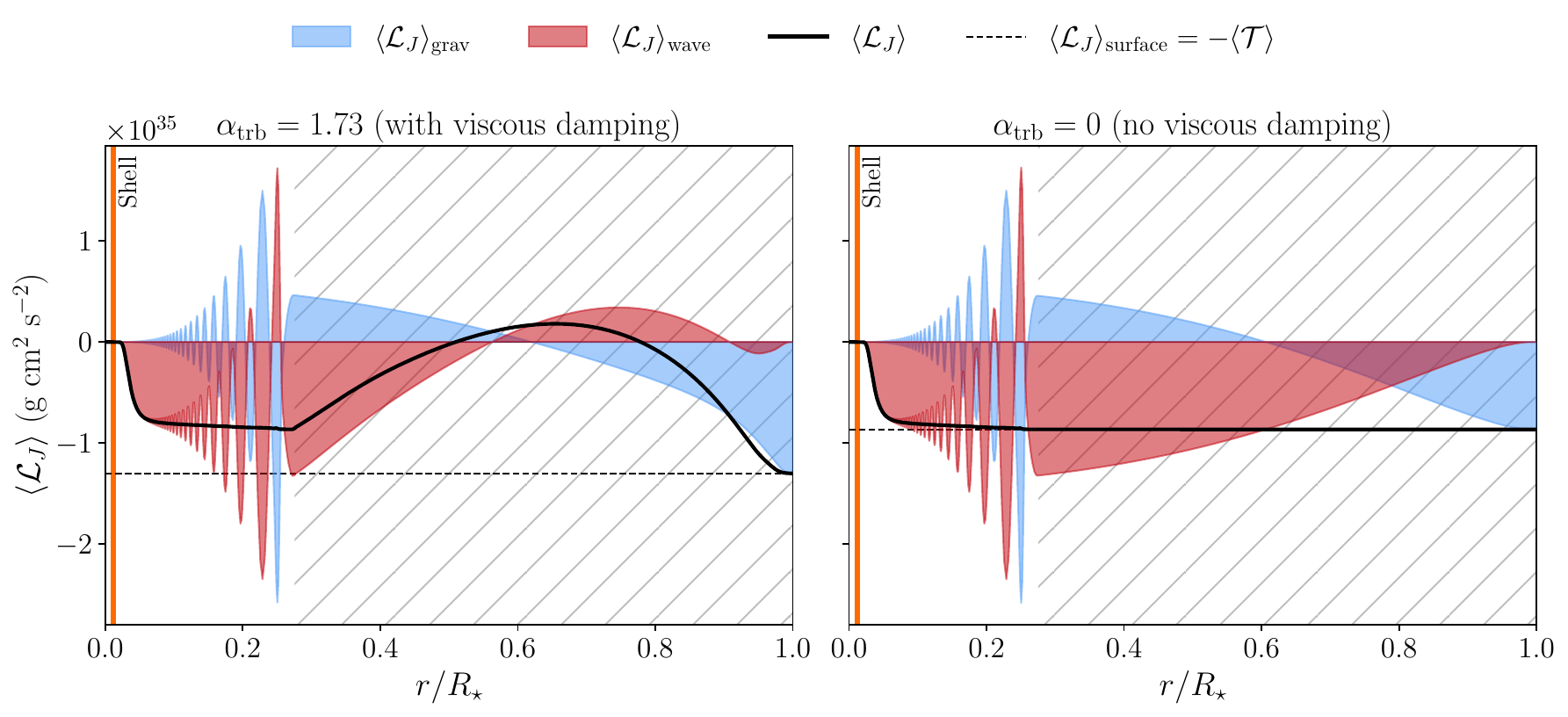}
    \caption{Angular momentum luminosity $\langle\mathcal{L}_{J}\rangle$ (black solid line) as a function of fractional stellar radius for the primary TOI-5882. To ensure consistency with Figures~\ref{fig:propagation} and \ref{fig:resonances}, these profiles are computed using the same stellar model, forced at the model-predicted orbital period of $P_{\rm orb} \approx 5.95$\,d. The right panel shows the \gyret calculation with only radiative damping, while the left panel includes both radiative and viscous damping. The blue shaded region represents the gravitational component of the luminosity $\langle\mathcal{L}_{J}\rangle_{\rm grav}$, and the red shaded region shows the wave component  $\langle \mathcal{L}_{J}\rangle_{\rm wave}$. The horizontal dashed line marks the surface value of $ \langle\mathcal{L}_{J}\rangle$, which is equal and opposite to the net torque $\langle\mathcal{T}\rangle$ on the star. The hatched region shows the convective zone of the star and the orange shaded zone represents the hydrogen burning shell.}
    \label{fig:J-luminosity}
\end{figure*}

In evolved stars, the localized deposition of angular momentum through the dissipation of IGWs near the hydrogen-burning shell can result in internal differential rotation \citep{fuller_angular_2014}. To explore the transport and deposition of angular momentum in TOI-5882, we also examine the angular momentum luminosity
\begin{equation}
\langle\mathcal{L}_{J}\rangle \equiv - \int_{0}^{r} \left\langle \frac{d\mathcal{T}}{dr'} \right\rangle \; dr',
\end{equation}
representing the orbit-averaged rate at which angular momentum flows outward through a spherical shell at radius $r$. Regions of the star where the radial gradient of $\langle\mathcal{L}_{J}\rangle$ is negative experience local deposition of angular momentum, equivalent to a positive torque on the star; and vice-versa for regions where the gradient is positive.

Figure~\ref{fig:J-luminosity} shows the radial profiles of $\langle\mathcal{L}_{J}\rangle$
for the same stellar model as Figures \ref{fig:propagation} and \ref{fig:resonances}, forced at the orbital period ($P_{\rm orb} \approx 5.95$ d) corresponding to the \mesa + \gyret tidal prescription at that evolutionary phase. The left and right panels show results with combined radiative and viscous damping and with pure radiative damping, respectively. We separate the total angular momentum luminosity (black solid line) into its constituent components: the red shaded region shows the wave luminosity $\langle\mathcal{L}_{J}\rangle_{\rm wave}$ arising from the first term on the right-hand side of Equation (\ref{eqn:differential_torque}), and the blue shaded region shows the gravitational luminosity $\langle\mathcal{L}_{J}\rangle_{\rm grav}$ arising from the second term. As physically required for conservative wave transport, $\langle\mathcal{L}_{J}\rangle_{\rm wave}$ approaches zero at the stellar surface; whereas $\langle\mathcal{L}_{J}\rangle_{\rm grav}$ and $\langle\mathcal{L}_{J}\rangle$ at the surface are together exactly equal and opposite to the net torque $\langle\mathcal{T}\rangle$ given by Equation (\ref{eqn:torque}).

In both panels of Figure~\ref{fig:J-luminosity}, the behavior seen in the radiative interior is very similar. Near the hydrogen-burning shell, the total angular momentum luminosity has a steep negative gradient, indicating the deposition of angular momentum corresponding to a positive torque. This deposition arises from strong radiative damping of the IGWs excited near the RCB. Further out in the radiative region, the damping is much weaker (although still present) and the gradient of $\langle\mathcal{L}_{J}\rangle$ becomes shallow.

The behavior in the convection zone is rather different in each panel. Near the stellar surface ($r/R_\star \gtrsim 0.85)$, the inclusion of viscous damping (left panel) results in a local steepening of the gravitational luminosity, compared to the case without viscous damping (right panel). As a result, the net torque on the star is increased by $\simeq 3 \times 10^{34}$ g cm$^{2}$ s$^{-2}$. Interestingly, this offset is the same torque predicted by the \citetalias{hut_tidal_1981} formalism for the equilibrium tide using this stellar model at the same forcing frequency (see Section~\ref{subsec:limitations_classicaltide} for further discussion).

Differences in the wave luminosity throughout the convection zone are also apparent. Without viscous damping due to turbulent viscosity, $\langle\mathcal{L}_{J}\rangle_{\rm wave}$ varies monotonically, and its gradient is equal and opposite to that of the gravitational luminosity, meaning that the total luminosity remains flat. When viscous damping is included, however, $\langle \mathcal{L}_{ J} \rangle_{\rm wave}$ grows rapidly from the RCB outward, until it becomes positive at $r/R_\star \approx 0.55$; beyond that point, it  continues to oscillate in sign but with an amplitude that decreases toward zero as the surface is approached.

\citet{goldreich_tidal_1989} provide the key to understanding the behaviors seen in Figure~\ref{fig:J-luminosity}. These authors argue that, in the absence of dissipation, the torques due to gravity and waves should be equal and opposite. This is precisely what is seen in the right-hand panel of the figure, where throughout the convective envelope (where there is no viscous or radiative damping) the gradients of $\langle\mathcal{L}_{J}\rangle_{\rm grav}$ and $\langle\mathcal{L}_{J}\rangle_{\rm wave}$ are equal and opposite. Conversely, in the left-hand panel, viscous damping causes a mismatch between these gradients, leading to a positive $\langle\mathcal{L}_{J}\rangle$ gradient and negative torque in the inner part of the convection zone ($0.28 \la r/R_\star \la 0.66$), and vice versa in the outer part ($r/R_\star \ga 0.66$).

The negative torque in the inner part of the convection zone might at first glance appear unphysical. However, as we intend to demonstrate in a follow-up paper (Townsend \& Narayan, in prep.), the inclusion of turbulent viscosity modifies $\langle \mathcal L_{J} \rangle$ through two distinct mechanisms. The first arises from viscous dissipation of the response excited by the tidal potential, and produces a local torque that is always positive. The second arises from viscous transport of angular momentum; in the case of Figure~\ref{fig:J-luminosity}, this transport extracts angular momentum from the inner part of the convection zone and deposits it in the outer part. Importantly, the transport is conservative, resulting in no net torque on the star. In the absence of other angular momentum redistribution processes (e.g., convection), the viscous transport mechanism would lead to the establishment of a differential rotation profile.

\section{Discussion} \label{section:discussion}

\subsection{Limitations of the Classical Tide Formalism} \label{subsec:limitations_classicaltide}

   \begin{figure}[htb!] 
        \includegraphics[width=0.96\linewidth]{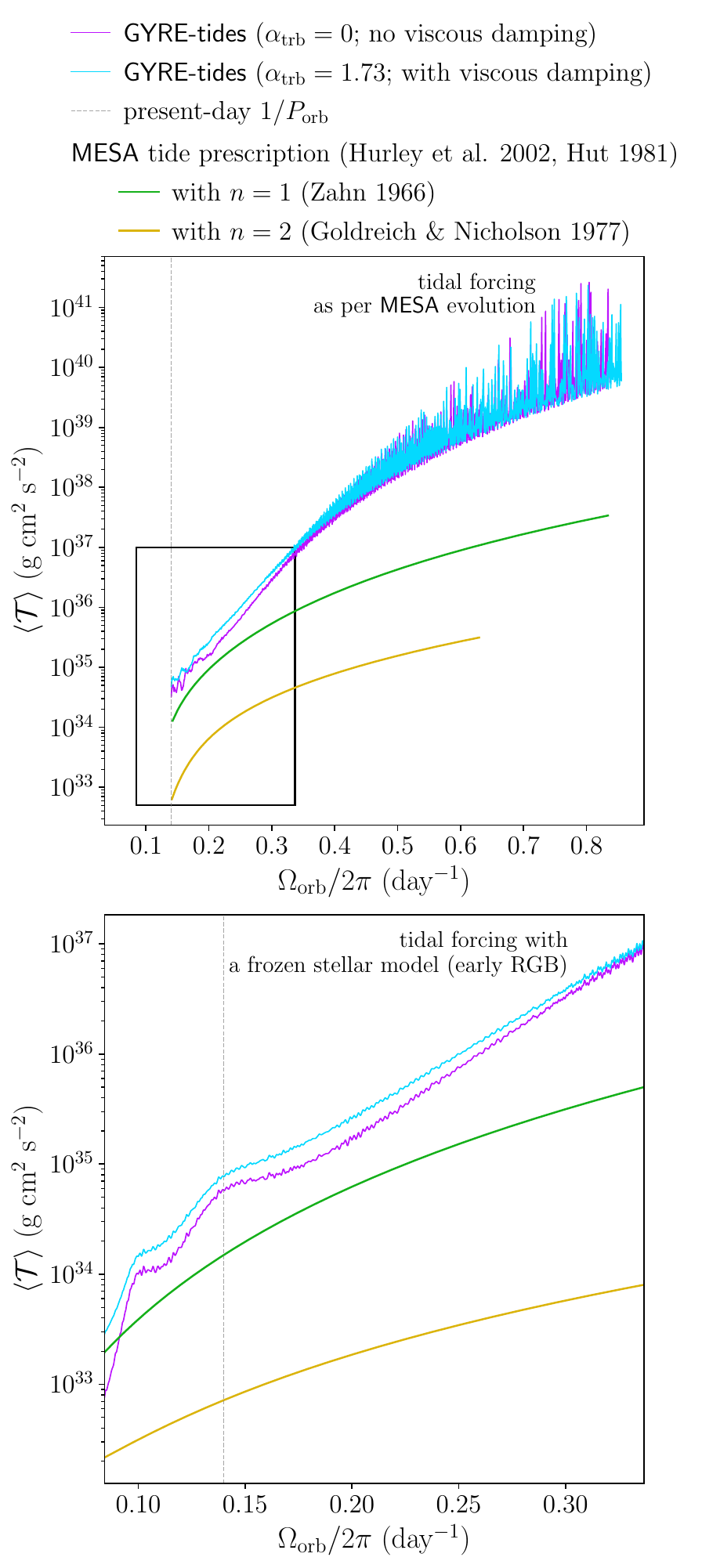}
        \caption{We show the total tidal torque as a function of the orbital forcing frequency, with the four tidal prescriptions. The top panel tracks the evolution of the torque as TOI-5882 evolves until RLOF, illustrating the onset of resonance crossings at higher forcing frequencies. The bottom panel isolates the angular momentum evolution in the early RGB phase by applying varying forcing frequencies to the same stellar model as Figures \ref{fig:propagation}, \ref{fig:resonances}, and \ref{fig:J-luminosity}, with the limits of the calculation indicated by the inset rectangle in the top panel. The vertical line shows the current period of the BD.} \label{fig:torque}
    \end{figure}

To contextualize the differences in the angular momentum evolution between the different tidal prescriptions, we must evaluate the regimes where the classical tidal prescriptions succeed and where they break down. To explore this, we investigate the evolution of tidal torque as a function of orbital forcing frequency, as illustrated by Figure~\ref{fig:torque}. To decouple the effects of a changing stellar structure from the purely frequency-dependent tidal response, we perform two complementary analyses. The top panel tracks the angular momentum exchange in the varying tidal prescriptions alongside the evolution of the primary star from its present-day orbital architecture. In contrast, the  bottom panel isolates the frequency dependence of the tides by applying a range of forcing frequencies to the same frozen stellar model as the other analyses in this Letter. 

At low forcing frequencies (where $\Omega_{\rm orb}$ is slightly greater than $\Omega_{\rm rot}$), the excitation and damping of IGWs are weak. Consequently, the tidal response is dominated by the turbulent viscous damping. As demonstrated by the lower panel of Figure~\ref{fig:torque}, in the low-frequency limit, the torque due to radiative damping rapidly decreases (magenta solid line) and the \gyret prescription with both radiative and viscous damping enabled (cyan solid line) closely parallels the equilibrium tide prediction. Specifically, the additional torque introduced by enabling viscous damping in our framework matches the torque predicted by the \citetalias{hurley_evolution_2002} prescription utilizing a linear ($n=1$) tidal reduction factor \citep{zahn_marees_1966}. The issue of the correct tidal reduction factor has often been considered as the `Achilles' heel of tidal theory' \citep{zahn_tidal_2008,duguid_convective_2020}. Rather than suggesting that the linear scaling is the correct one, this alignment demonstrates the consistency between the equilibrium tide framework and the \gyret framework \citep{sun_numerical_2025,sun_tides_2026}. 

To explicitly confirm this, we performed a supplementary \gyret calculation with the same stellar model as shown in the figures in this Letter, employing the static tide approximation (by setting the threshold on the corotating frequency for a partial tide, \texttt{omega\_c\_thresh}, arbitrarily high). 
The resulting tidal response shows that the displacement eigenfunction, $\tilde{\xi}_r$, closely matches its non-static counterpart in the convective zone, indicating that the full \gyret response captures equilibrium tide-like behavior in the appropriate limits. 

However, as the system evolves to higher frequencies (including the present-day forcing frequency of the TOI-5882b system), the equilibrium tide model begins to fail. Irrespective of the choice of tidal reduction factor, the \citetalias{hurley_evolution_2002} formalism underpredicts the total torque by several orders of magnitude as this model assumes only tidal dissipation in the convective envelope. The necessity of detailed tidal modeling in radiative regions has also been recently highlighted by \citet{esseldeurs_competition_2026}, who investigated the competition between stochastically excited IGWs and tidally forced IGWs. By modeling the energy and angular momentum luminosities of both wave types, they demonstrated that while stochastic excitation dominates for Jupiter-mass planets, tidally excited IGWs carry the larger angular momentum luminosity for massive companions that are on orbits shorter than a few days. 

As a massive BD on a short-period orbit, TOI-5882b bridges this gap as it is in the regime where tidal IGW excitation becomes important. By partitioning the tidal response into two distinct processes---an ``equilibrium tide'' or a ``dynamical tide''---the \citetalias{hut_tidal_1981} and \citetalias{hurley_evolution_2002} formalism fundamentally misses the star's unified response to the tidal potential. Therefore, rather than adhering to this dichotomy, an accurate model of tidal dissipation must reframe the interaction entirely around the dissipation mechanisms governing the tidal response. 

\subsection{Radiatively and Viscously Damped Tides} \label{subsec:new_tide_formalism}

The case for moving beyond this dichotomy is further reinforced by the inconsistent usage of the term ``dynamical tide'' across the literature. In its original formulation, the dynamical tide referred specifically to the radiative damping of tidally-excited $g$-modes in the radiative envelopes of early-type stars, with the equilibrium tide reserved for the turbulent dissipation in the convective envelopes of late-type stars \citep{zahn_dynamical_1975,zahn_tidal_1977}. The \citetalias{hurley_evolution_2002} prescription preserves this separation, evaluating the dynamical tide only in stellar envelopes through a fit to the $E_2$ coupling coefficient of \citet{zahn_dynamical_1975}. Despite the narrow regime of its intended validity, this prescription is routinely applied as the default tidal treatment in binary evolution and population synthesis codes, where its underlying assumptions break down \citep{kushnir_dynamical_2017,mirouh_detailed_2023,sciarini_dynamical_2024}. 

A parallel literature has long recognized that radiative damping of IGWs can not only occur in the envelope but also in the radiative interiors. \citet{goodman_dynamical_1998} showed that resonant $g$-mode excitation in the cores of solar-type stars can circularize binaries on timescales comparable to tidal dissipation in the envelope, while \citet{terquem_tidal_1998} modeled the tidal response of a solar-type star with viscous and radiative damping simultaneously. \citet{ogilvie_tidal_2014} subsequently reframed this dichotomy by identifying the equilibrium tide with the non-wavelike component of the tidal response and the dynamical tide with the wavelike response, irrespective of stellar geometry. A separate line of work treats the dynamical tide in the context of a resonance, in which the tidally excited IGWs can lock to the orbital evolution \citep[e.g.,][]{witte_orbital_2002,fuller_angular_2014,ma_orbital_2021,bryan_coevolution_2024}. These three usages are physically related but not interchangeable, and the inherited language has obscured the underlying physics. 

These limitations motivate a different organizing principle. Rather than partitioning the tidal response by its assumed character (equilibrium vs. dynamical), we propose that the response be partitioned by its underlying dissipation mechanism: radiatively and viscously damped tides. Both mechanisms operate simultaneously throughout any star, with their relative importance set by the local stellar structure. In stably stratified regions regions where IGWs propagate, the dissipation is a consequence of radiative diffusion. In convective regions, where IGWs become evanescent and the tidal response resembles the equilibrium tide, the dissipation primarily arises from turbulent viscosity. In both cases, the dissipation results in the local deposition of energy and angular momentum extacted from the orbit.

Our \mesa + \gyret framework lends itself to this reframing. By solving the linearized nonadiabatic forced oscillation equations augmented with a radial viscous force, both dissipation mechanisms contribute to the tidal response at the level of local fluid perturbations rather than separate global prescriptions. Resonant mode excitations emerge naturally when the IGWs can form a standing wave with frequencies near integer multiples of the orbital forcing frequency, as seen by the resonance crossings in the upper panel of Figure~\ref{fig:torque}. In the low-frequency limit, this formulation reproduces the \citetalias{hut_tidal_1981} and \citetalias{hurley_evolution_2002} equilibrium tide prescription, recovering the historical dichotomy while remaining valid across a range of stellar evolution phases. 

\subsection{Future Work \& Implications}

To fully model the eventual fate of TOI-5882b, we will need to investigate the processes beyond RLOF. To accurately model unstable mass transfer and the subsequent common envelope evolution phase will require a fully hydrodynamic treatment. Thus, future work to understand the evolution of this system will involve such simulations initialized at RLOF to allow for a more physically complete assessment of the engulfment process.  While the almost circular orbit of TOI-5882b allowed us to treat the change in eccentricity as negligible, future applications of this framework to high-eccentricity migration or eccentric binaries will require extending the \mesa binary interface to accept externally computed $\dot{e}$ rates. 

The coupled tidal and stellar evolution framework we present here also has broad implications for a diverse range of close-in systems. For instance, our framework can be used to robustly model the migration and survival of hot Jupiters discovered orbiting evolved hosts \citep{grunblatt_tess_2022}. By incorporating radiative damping, our framework provides a physical basis for the enhanced tidal dissipation rates potentially needed to explain existing discrepancies in binary population statistics. Specifically, by applying this self-consistent modeling to rederive the RLOF stability criteria determined by the \citet{rappaport_relation_1995} relation, we can more precisely determine the threshold between stable and unstable RLOF, especially for anomalous systems like the massive white dwarf identified by \citet{motherway_not-so-compact_2026}. This could also help explain other anomalous systems, such as the missing population of binaries in NGC 188 with orbital periods longer than the limit predicted by RLOF-driven evolution \citep{narayan_wiyn_2026}, thereby helping to map the tidal pathways that could impact the formation of compact remnants and short-period binaries.

\begin{acknowledgements}
We are grateful to the anonymous reviewer for providing thoughtful comments that improved the scientific analyses presented in this paper. We gratefully acknowledge support from the \textit{Peter Livingston Scholars Program}, whose commitment to undergraduate research has been instrumental in advancing this work.
Support for this research was provided by the Office of the Vice Chancellor for Research and Graduate Education at the University of Wisconsin--Madison with funding from the Wisconsin Alumni Research Foundation. R.H.D.T. acknowledges support from NASA ATP award 80NSSC24K0895 and NSF AAG award 2407636. R.S.N. thanks Meng Sun, Fred Rasio, and Adam Distler, Vincent Vanlaer, and Jarett Rosenberg for valuable discussions that helped shape the science in this work. 

This work was conducted at the University of Wisconsin-Madison, which is located on occupied ancestral land of the Ho-Chunk people, a place their nation has called Teejop since time immemorial. In an 1832 treaty, the Ho-Chunk were forced to cede this territory. The university was founded on and funded through this seized land; this legacy enabled the science presented here. 
\end{acknowledgements}

\software{\textsf{Astropy} \citep{astropy_collaboration_astropy_2013,astropy_collaboration_astropy_2018,astropy_collaboration_astropy_2022}, \textsf{GYRE} \citep{townsend_gyre_2013,townsend_angular_2018,sun_gyre_tides_2023}, \textsf{MESA} \citep{paxton_modules_2011,paxton_modules_2013,paxton_modules_2015,paxton_modules_2018,paxton_modules_2019,jermyn_modules_2023}, \textsf{MESAsdk} \citep{townsend_mesa_2026}, \textsf{Numpy} \citep{harris_array_2020}, \textsf{Scipy} \citep{virtanen_scipy_2020}, \textsf{pandas} \citep{mckinney_data_2010}.}

\bibliography{references}{}

@article{goldreich_tidal_1989,
	title = {Tidal {Friction} in {Early}-{Type} {Stars}},
	volume = {342},
	issn = {0004-637X},
	url = {https://ui.adsabs.harvard.edu/abs/1989ApJ...342.1079G},
	doi = {10.1086/167665},
	urldate = {2026-05-23},
	journal = {The Astrophysical Journal},
	publisher = {IOP},
	author = {Goldreich, Peter and Nicholson, Philip D.},
	month = jul,
	year = {1989},
	note = {ADS Bibcode: 1989ApJ...342.1079G},
	pages = {1079},
}

@misc{townsend_mesa_2026,
	title = {{MESA} {SDK} for {Mac} {OS}},
	url = {https://zenodo.org/records/19364818},
	doi = {10.5281/zenodo.19364818},
	urldate = {2026-05-03},
	publisher = {Zenodo},
	author = {Townsend, Richard},
	month = mar,
	year = {2026},
}

@phdthesis{bonsor_post-main_2011,
	type = {Ph.{D}. thesis},
	title = {Post-main sequence evolution of planetary systems},
	url = {https://ui.adsabs.harvard.edu/abs/2011PhDT.......445B},
	urldate = {2026-04-17},
	school = {University of Cambridge},
	author = {Bonsor, Amy Hannah Clay},
	month = jan,
	year = {2011},
	note = {ADS Bibcode: 2011PhDT.......445B},
}

@book{aerts_asteroseismology_2010,
	title = {Asteroseismology},
	url = {https://ui.adsabs.harvard.edu/abs/2010aste.book.....A},
	doi = {10.1007/978-1-4020-5803-5},
	urldate = {2026-04-22},
	publisher = {Springer},
	author = {Aerts, Conny and Christensen-Dalsgaard, Jørgen and Kurtz, Donald W.},
	month = jan,
	year = {2010},
	note = {Publication Title: Asteroseismology
ADS Bibcode: 2010aste.book.....A},
}

@article{mckinney_data_2010,
	title = {Data {Structures} for {Statistical} {Computing} in {Python}},
	url = {https://proceedings.scipy.org/articles/Majora-92bf1922-00a},
	doi = {10.25080/Majora-92bf1922-00a},
	language = {en},
	urldate = {2025-08-29},
	journal = {scipy},
	author = {McKinney, Wes},
	month = may,
	year = {2010},
}

@article{virtanen_scipy_2020,
	title = {{SciPy} 1.0: fundamental algorithms for scientific computing in {Python}},
	volume = {17},
	copyright = {2020 The Author(s)},
	issn = {1548-7105},
	shorttitle = {{SciPy} 1.0},
	url = {https://www.nature.com/articles/s41592-019-0686-2},
	doi = {10.1038/s41592-019-0686-2},
	language = {en},
	number = {3},
	urldate = {2025-08-29},
	journal = {Nature Methods},
	publisher = {Nature Publishing Group},
	author = {Virtanen, Pauli and Gommers, Ralf and Oliphant, Travis E. and Haberland, Matt and Reddy, Tyler and Cournapeau, David and Burovski, Evgeni and Peterson, Pearu and Weckesser, Warren and Bright, Jonathan and van der Walt, Stéfan J. and Brett, Matthew and Wilson, Joshua and Millman, K. Jarrod and Mayorov, Nikolay and Nelson, Andrew R. J. and Jones, Eric and Kern, Robert and Larson, Eric and Carey, C. J. and Polat, İlhan and Feng, Yu and Moore, Eric W. and VanderPlas, Jake and Laxalde, Denis and Perktold, Josef and Cimrman, Robert and Henriksen, Ian and Quintero, E. A. and Harris, Charles R. and Archibald, Anne M. and Ribeiro, Antônio H. and Pedregosa, Fabian and van Mulbregt, Paul},
	month = mar,
	year = {2020},
	pages = {261--272},
}

@article{harris_array_2020,
	title = {Array programming with {NumPy}},
	volume = {585},
	copyright = {2020 The Author(s)},
	issn = {1476-4687},
	url = {https://www.nature.com/articles/s41586-020-2649-2},
	doi = {10.1038/s41586-020-2649-2},
	language = {en},
	number = {7825},
	urldate = {2025-08-29},
	journal = {Nature},
	publisher = {Nature Publishing Group},
	author = {Harris, Charles R. and Millman, K. Jarrod and van der Walt, Stéfan J. and Gommers, Ralf and Virtanen, Pauli and Cournapeau, David and Wieser, Eric and Taylor, Julian and Berg, Sebastian and Smith, Nathaniel J. and Kern, Robert and Picus, Matti and Hoyer, Stephan and van Kerkwijk, Marten H. and Brett, Matthew and Haldane, Allan and del Río, Jaime Fernández and Wiebe, Mark and Peterson, Pearu and Gérard-Marchant, Pierre and Sheppard, Kevin and Reddy, Tyler and Weckesser, Warren and Abbasi, Hameer and Gohlke, Christoph and Oliphant, Travis E.},
	month = sep,
	year = {2020},
	pages = {357--362},
}

@article{astropy_collaboration_astropy_2022,
	title = {The {Astropy} {Project}: {Sustaining} and {Growing} a {Community}-oriented {Open}-source {Project} and the {Latest} {Major} {Release} (v5.0) of the {Core} {Package}},
	volume = {935},
	issn = {0004-637X},
	shorttitle = {The {Astropy} {Project}},
	url = {https://ui.adsabs.harvard.edu/abs/2022ApJ...935..167A},
	doi = {10.3847/1538-4357/ac7c74},
	urldate = {2025-08-29},
	journal = {The Astrophysical Journal},
	author = {{Astropy Collaboration} and Price-Whelan, Adrian M. and Lim, Pey Lian and Earl, Nicholas and Starkman, Nathaniel and Bradley, Larry and Shupe, David L. and Patil, Aarya A. and Corrales, Lia and Brasseur, C. E. and Nöthe, Maximilian and Donath, Axel and Tollerud, Erik and Morris, Brett M. and Ginsburg, Adam and Vaher, Eero and Weaver, Benjamin A. and Tocknell, James and Jamieson, William and van Kerkwijk, Marten H. and Robitaille, Thomas P. and Merry, Bruce and Bachetti, Matteo and Günther, H. Moritz and Aldcroft, Thomas L. and Alvarado-Montes, Jaime A. and Archibald, Anne M. and Bódi, Attila and Bapat, Shreyas and Barentsen, Geert and Bazán, Juanjo and Biswas, Manish and Boquien, Médéric and Burke, D. J. and Cara, Daria and Cara, Mihai and Conroy, Kyle E. and Conseil, Simon and Craig, Matthew W. and Cross, Robert M. and Cruz, Kelle L. and D'Eugenio, Francesco and Dencheva, Nadia and Devillepoix, Hadrien A. R. and Dietrich, Jörg P. and Eigenbrot, Arthur Davis and Erben, Thomas and Ferreira, Leonardo and Foreman-Mackey, Daniel and Fox, Ryan and Freij, Nabil and Garg, Suyog and Geda, Robel and Glattly, Lauren and Gondhalekar, Yash and Gordon, Karl D. and Grant, David and Greenfield, Perry and Groener, Austen M. and Guest, Steve and Gurovich, Sebastian and Handberg, Rasmus and Hart, Akeem and Hatfield-Dodds, Zac and Homeier, Derek and Hosseinzadeh, Griffin and Jenness, Tim and Jones, Craig K. and Joseph, Prajwel and Kalmbach, J. Bryce and Karamehmetoglu, Emir and Kałuszyński, Mikołaj and Kelley, Michael S. P. and Kern, Nicholas and Kerzendorf, Wolfgang E. and Koch, Eric W. and Kulumani, Shankar and Lee, Antony and Ly, Chun and Ma, Zhiyuan and MacBride, Conor and Maljaars, Jakob M. and Muna, Demitri and Murphy, N. A. and Norman, Henrik and O'Steen, Richard and Oman, Kyle A. and Pacifici, Camilla and Pascual, Sergio and Pascual-Granado, J. and Patil, Rohit R. and Perren, Gabriel I. and Pickering, Timothy E. and Rastogi, Tanuj and Roulston, Benjamin R. and Ryan, Daniel F. and Rykoff, Eli S. and Sabater, Jose and Sakurikar, Parikshit and Salgado, Jesús and Sanghi, Aniket and Saunders, Nicholas and Savchenko, Volodymyr and Schwardt, Ludwig and Seifert-Eckert, Michael and Shih, Albert Y. and Jain, Anany Shrey and Shukla, Gyanendra and Sick, Jonathan and Simpson, Chris and Singanamalla, Sudheesh and Singer, Leo P. and Singhal, Jaladh and Sinha, Manodeep and Sipőcz, Brigitta M. and Spitler, Lee R. and Stansby, David and Streicher, Ole and Šumak, Jani and Swinbank, John D. and Taranu, Dan S. and Tewary, Nikita and Tremblay, Grant R. and de Val-Borro, Miguel and Van Kooten, Samuel J. and Vasović, Zlatan and Verma, Shresth and de Miranda Cardoso, José Vinícius and Williams, Peter K. G. and Wilson, Tom J. and Winkel, Benjamin and Wood-Vasey, W. M. and Xue, Rui and Yoachim, Peter and Zhang, Chen and Zonca, Andrea and {Astropy Project Contributors}},
	month = aug,
	year = {2022},
	note = {ADS Bibcode: 2022ApJ...935..167A},
	pages = {167},
}

@article{astropy_collaboration_astropy_2018,
	title = {The {Astropy} {Project}: {Building} an {Open}-science {Project} and {Status} of the v2.0 {Core} {Package}},
	volume = {156},
	issn = {0004-6256},
	shorttitle = {The {Astropy} {Project}},
	url = {https://ui.adsabs.harvard.edu/abs/2018AJ....156..123A},
	doi = {10.3847/1538-3881/aabc4f},
	urldate = {2025-08-29},
	journal = {The Astronomical Journal},
	author = {{Astropy Collaboration} and Price-Whelan, A. M. and Sipőcz, B. M. and Günther, H. M. and Lim, P. L. and Crawford, S. M. and Conseil, S. and Shupe, D. L. and Craig, M. W. and Dencheva, N. and Ginsburg, A. and VanderPlas, J. T. and Bradley, L. D. and Pérez-Suárez, D. and de Val-Borro, M. and Aldcroft, T. L. and Cruz, K. L. and Robitaille, T. P. and Tollerud, E. J. and Ardelean, C. and Babej, T. and Bach, Y. P. and Bachetti, M. and Bakanov, A. V. and Bamford, S. P. and Barentsen, G. and Barmby, P. and Baumbach, A. and Berry, K. L. and Biscani, F. and Boquien, M. and Bostroem, K. A. and Bouma, L. G. and Brammer, G. B. and Bray, E. M. and Breytenbach, H. and Buddelmeijer, H. and Burke, D. J. and Calderone, G. and Cano Rodríguez, J. L. and Cara, M. and Cardoso, J. V. M. and Cheedella, S. and Copin, Y. and Corrales, L. and Crichton, D. and D'Avella, D. and Deil, C. and Depagne, É. and Dietrich, J. P. and Donath, A. and Droettboom, M. and Earl, N. and Erben, T. and Fabbro, S. and Ferreira, L. A. and Finethy, T. and Fox, R. T. and Garrison, L. H. and Gibbons, S. L. J. and Goldstein, D. A. and Gommers, R. and Greco, J. P. and Greenfield, P. and Groener, A. M. and Grollier, F. and Hagen, A. and Hirst, P. and Homeier, D. and Horton, A. J. and Hosseinzadeh, G. and Hu, L. and Hunkeler, J. S. and Ivezić, Ž. and Jain, A. and Jenness, T. and Kanarek, G. and Kendrew, S. and Kern, N. S. and Kerzendorf, W. E. and Khvalko, A. and King, J. and Kirkby, D. and Kulkarni, A. M. and Kumar, A. and Lee, A. and Lenz, D. and Littlefair, S. P. and Ma, Z. and Macleod, D. M. and Mastropietro, M. and McCully, C. and Montagnac, S. and Morris, B. M. and Mueller, M. and Mumford, S. J. and Muna, D. and Murphy, N. A. and Nelson, S. and Nguyen, G. H. and Ninan, J. P. and Nöthe, M. and Ogaz, S. and Oh, S. and Parejko, J. K. and Parley, N. and Pascual, S. and Patil, R. and Patil, A. A. and Plunkett, A. L. and Prochaska, J. X. and Rastogi, T. and Reddy Janga, V. and Sabater, J. and Sakurikar, P. and Seifert, M. and Sherbert, L. E. and Sherwood-Taylor, H. and Shih, A. Y. and Sick, J. and Silbiger, M. T. and Singanamalla, S. and Singer, L. P. and Sladen, P. H. and Sooley, K. A. and Sornarajah, S. and Streicher, O. and Teuben, P. and Thomas, S. W. and Tremblay, G. R. and Turner, J. E. H. and Terrón, V. and van Kerkwijk, M. H. and de la Vega, A. and Watkins, L. L. and Weaver, B. A. and Whitmore, J. B. and Woillez, J. and Zabalza, V. and {Astropy Contributors}},
	month = sep,
	year = {2018},
	note = {ADS Bibcode: 2018AJ....156..123A},
	pages = {123},
}

@article{astropy_collaboration_astropy_2013,
	title = {Astropy: {A} community {Python} package for astronomy},
	volume = {558},
	issn = {0004-6361},
	shorttitle = {Astropy},
	url = {https://ui.adsabs.harvard.edu/abs/2013A&A...558A..33A},
	doi = {10.1051/0004-6361/201322068},
	urldate = {2025-08-29},
	journal = {Astronomy and Astrophysics},
	author = {{Astropy Collaboration} and Robitaille, Thomas P. and Tollerud, Erik J. and Greenfield, Perry and Droettboom, Michael and Bray, Erik and Aldcroft, Tom and Davis, Matt and Ginsburg, Adam and Price-Whelan, Adrian M. and Kerzendorf, Wolfgang E. and Conley, Alexander and Crighton, Neil and Barbary, Kyle and Muna, Demitri and Ferguson, Henry and Grollier, Frédéric and Parikh, Madhura M. and Nair, Prasanth H. and Unther, Hans M. and Deil, Christoph and Woillez, Julien and Conseil, Simon and Kramer, Roban and Turner, James E. H. and Singer, Leo and Fox, Ryan and Weaver, Benjamin A. and Zabalza, Victor and Edwards, Zachary I. and Azalee Bostroem, K. and Burke, D. J. and Casey, Andrew R. and Crawford, Steven M. and Dencheva, Nadia and Ely, Justin and Jenness, Tim and Labrie, Kathleen and Lim, Pey Lian and Pierfederici, Francesco and Pontzen, Andrew and Ptak, Andy and Refsdal, Brian and Servillat, Mathieu and Streicher, Ole},
	month = oct,
	year = {2013},
	note = {ADS Bibcode: 2013A\&A...558A..33A},
	pages = {A33},
}

@article{rappaport_relation_1995,
	title = {The relation between white dwarf mass and orbital period in wide binary radio pulsars},
	volume = {273},
	issn = {0035-8711},
	url = {https://ui.adsabs.harvard.edu/abs/1995MNRAS.273..731R},
	doi = {10.1093/mnras/273.3.731},
	urldate = {2026-04-25},
	journal = {Monthly Notices of the Royal Astronomical Society},
	publisher = {OUP},
	author = {Rappaport, S. and Podsiadlowski, Ph. and Joss, P. C. and Di Stefano, R. and Han, Z.},
	month = apr,
	year = {1995},
	note = {ADS Bibcode: 1995MNRAS.273..731R},
	pages = {731--741},
}

@article{motherway_not-so-compact_2026,
	title = {A {Not}-so-compact {Companion}: {Massive}, {Oversized} {White} {Dwarf} in a {Post}-common-envelope {Eclipsing} {Binary}},
	volume = {171},
	issn = {0004-6256},
	shorttitle = {A {Not}-so-compact {Companion}},
	url = {https://ui.adsabs.harvard.edu/abs/2026AJ....171..159M},
	doi = {10.3847/1538-3881/ae3b42},
	urldate = {2026-04-25},
	journal = {The Astronomical Journal},
	publisher = {IOP},
	author = {Motherway, Erin and Linck, Evan and Mathieu, Robert D. and Dixon, Don and Stassun, Keivan G. and Breivik, Katelyn and Majewski, Steve and Pols, Onno R.},
	month = mar,
	year = {2026},
	note = {ADS Bibcode: 2026AJ....171..159M},
	pages = {159},
}

@article{grunblatt_tess_2022,
	title = {{TESS} {Giants} {Transiting} {Giants}. {II}. {The} {Hottest} {Jupiters} {Orbiting} {Evolved} {Stars}},
	volume = {163},
	issn = {1538-3881},
	url = {https://doi.org/10.3847/1538-3881/ac4972},
	doi = {10.3847/1538-3881/ac4972},
	language = {en},
	number = {3},
	urldate = {2026-04-25},
	journal = {The Astronomical Journal},
	publisher = {The American Astronomical Society},
	author = {Grunblatt, Samuel K. and Saunders, Nicholas and Sun, Meng and Chontos, Ashley and Soares-Furtado, Melinda and Eisner, Nora and Pereira, Filipe and Komacek, Thaddeus and Huber, Daniel and Collins, Karen and Wang, Gavin and Stockdale, Chris and Quinn, Samuel N. and Tronsgaard, Rene and Zhou, George and Nowak, Grzegorz and Deeg, Hans J. and Ciardi, David R. and Boyle, Andrew and Rice, Malena and Dai, Fei and Blunt, Sarah and Van Zandt, Judah and Beard, Corey and Akana Murphy, Joseph M. and Dalba, Paul A. and Lubin, Jack and Polanski, Alex and Brinkman, Casey Lynn and Howard, Andrew W. and Buchhave, Lars A. and Angus, Ruth and Ricker, George R. and Jenkins, Jon M. and Wohler, Bill and Goeke, Robert F. and Levine, Alan M. and Colon, Knicole D. and Huang, Chelsea X. and Kunimoto, Michelle and Shporer, Avi and Latham, David W. and Seager, Sara and Vanderspek, Roland K. and Winn, Joshua N.},
	month = feb,
	year = {2022},
	pages = {120},
}

@article{bryan_coevolution_2024,
	title = {The coevolution of migrating planets and their pulsating stars through episodic resonance locking},
	volume = {8},
	issn = {2397-3366},
	url = {https://ui.adsabs.harvard.edu/abs/2024NatAs...8.1387B},
	doi = {10.1038/s41550-024-02351-3},
	urldate = {2026-04-25},
	journal = {Nature Astronomy},
	author = {Bryan, Jared and de Wit, Julien and Sun, Meng and de Beurs, Zoë L. and Townsend, Richard H. D.},
	month = nov,
	year = {2024},
	note = {ADS Bibcode: 2024NatAs...8.1387B},
	pages = {1387--1398},
}

@article{kushnir_dynamical_2017,
	title = {Dynamical {Tides} {Reexpressed}},
	volume = {467},
	issn = {0035-8711, 1365-2966},
	url = {http://arxiv.org/abs/1605.03810},
	doi = {10.1093/mnras/stx255},
	number = {2},
	urldate = {2026-04-25},
	journal = {Monthly Notices of the Royal Astronomical Society},
	author = {Kushnir, Doron and Zaldarriaga, Matias and Kollmeier, Juna A. and Waldman, Roni},
	month = may,
	year = {2017},
	note = {arXiv:1605.03810 [astro-ph]},
	pages = {2146--2149},
}

@article{sciarini_dynamical_2024,
	title = {Dynamical tides in binaries: {Inconsistencies} in the implementation of {Zahn}'s prescription},
	volume = {681},
	issn = {0004-6361},
	shorttitle = {Dynamical tides in binaries},
	url = {https://ui.adsabs.harvard.edu/abs/2024A&A...681L...1S},
	doi = {10.1051/0004-6361/202348424},
	urldate = {2026-04-25},
	journal = {Astronomy and Astrophysics},
	publisher = {EDP},
	author = {Sciarini, Luca and Ekström, Sylvia and Eggenberger, Patrick and Meynet, Georges and Fragos, Tassos and Song, Han Feng},
	month = jan,
	year = {2024},
	note = {ADS Bibcode: 2024A\&A...681L...1S},
	pages = {L1},
}

@article{sun_tides_2026,
	title = {Tides in {Massive} {Binaries}: {Numerical} {Solutions} and {Semianalytical} {Comparisons}},
	volume = {998},
	issn = {0004-637X},
	shorttitle = {Tides in {Massive} {Binaries}},
	url = {https://doi.org/10.3847/1538-4357/ae2dff},
	doi = {10.3847/1538-4357/ae2dff},
	language = {en},
	number = {1},
	urldate = {2026-04-24},
	journal = {The Astrophysical Journal},
	publisher = {The American Astronomical Society},
	author = {Sun, Meng and Xia, Hongbo and Gossage, Seth and Kalogera, Vicky and Liu, Jifeng and Rocha, Kyle Akira and Townsend, R. H. D. and Zapartas, Emmanouil},
	month = feb,
	year = {2026},
	pages = {32},
}

@article{ivanov_unified_2013,
	title = {A unified normal mode approach to dynamic tides and its application to rotating {Sun}-like stars},
	volume = {432},
	issn = {0035-8711},
	url = {https://ui.adsabs.harvard.edu/abs/2013MNRAS.432.2339I},
	doi = {10.1093/mnras/stt595},
	urldate = {2026-04-22},
	journal = {Monthly Notices of the Royal Astronomical Society},
	publisher = {OUP},
	author = {Ivanov, P. B. and Papaloizou, J. C. B. and Chernov, S. V.},
	month = jul,
	year = {2013},
	note = {ADS Bibcode: 2013MNRAS.432.2339I},
	pages = {2339--2365},
}

@article{gallet_tidal_2017,
	title = {Tidal dissipation in rotating low-mass stars and implications for the orbital evolution of close-in planets. {I}. {From} the {PMS} to the {RGB} at solar metallicity},
	volume = {604},
	issn = {0004-6361},
	url = {https://ui.adsabs.harvard.edu/abs/2017A&A...604A.112G},
	doi = {10.1051/0004-6361/201730661},
	urldate = {2026-04-22},
	journal = {Astronomy and Astrophysics},
	publisher = {EDP},
	author = {Gallet, F. and Bolmont, E. and Mathis, S. and Charbonnel, C. and Amard, L.},
	month = aug,
	year = {2017},
	note = {ADS Bibcode: 2017A\&A...604A.112G},
	pages = {A112},
}

@article{bolmont_effect_2016,
	title = {Effect of the rotation and tidal dissipation history of stars on the evolution of close-in planets},
	volume = {126},
	issn = {0923-2958},
	url = {https://ui.adsabs.harvard.edu/abs/2016CeMDA.126..275B},
	doi = {10.1007/s10569-016-9690-3},
	urldate = {2026-04-22},
	journal = {Celestial Mechanics and Dynamical Astronomy},
	publisher = {Springer},
	author = {Bolmont, Emeline and Mathis, Stéphane},
	month = nov,
	year = {2016},
	note = {ADS Bibcode: 2016CeMDA.126..275B},
	pages = {275--296},
}

@misc{zahn_angular_1997,
	title = {Angular momentum transport by internal waves in the solar interior.},
	url = {https://ui.adsabs.harvard.edu/abs/1997A&A...322..320Z},
	doi = {10.48550/arXiv.astro-ph/9611189},
	urldate = {2026-04-22},
	publisher = {arXiv},
	author = {Zahn, J.-P. and Talon, S. and Matias, J.},
	month = jun,
	year = {1997},
	note = {ISSN: 0004-6361
Volume: 322
ADS Bibcode: 1997A\&A...322..320Z},
}

@book{wu_hansen_2024,
	series = {Springer {Aerospace} {Technology}},
	title = {Hansen {Coefficients} in {Satellite} {Orbital} {Dynamics}},
	isbn = {978-981-97-0456-9},
	url = {https://books.google.com/books?id=Ss0DEQAAQBAJ},
	publisher = {Springer Nature Singapore},
	author = {Wu, L. and Zhang, M.},
	year = {2024},
}

@article{hon_asteroseismic_2020,
	title = {Asteroseismic inference of subgiant evolutionary parameters with deep learning},
	volume = {499},
	issn = {0035-8711},
	url = {https://ui.adsabs.harvard.edu/abs/2020MNRAS.499.2445H},
	doi = {10.1093/mnras/staa2853},
	urldate = {2026-04-20},
	journal = {Monthly Notices of the Royal Astronomical Society},
	publisher = {OUP},
	author = {Hon, Marc and Bellinger, Earl P. and Hekker, Saskia and Stello, Dennis and Kuszlewicz, James S.},
	month = dec,
	year = {2020},
	note = {ADS Bibcode: 2020MNRAS.499.2445H},
	pages = {2445--2461},
}

@article{fuller_angular_2014,
	title = {{ANGULAR} {MOMENTUM} {TRANSPORT} {VIA} {INTERNAL} {GRAVITY} {WAVES} {IN} {EVOLVING} {STARS}},
	volume = {796},
	issn = {0004-637X},
	url = {https://doi.org/10.1088/0004-637X/796/1/17},
	doi = {10.1088/0004-637X/796/1/17},
	language = {en},
	number = {1},
	urldate = {2026-04-18},
	journal = {The Astrophysical Journal},
	publisher = {The American Astronomical Society},
	author = {Fuller, Jim and Lecoanet, Daniel and Cantiello, Matteo and Brown, Ben},
	month = oct,
	year = {2014},
	pages = {17},
}

@article{li_can_2025,
	title = {Can tidal evolution lead to close-in planetary bodies around white dwarfs - {I}. {Orbital} period distribution},
	volume = {537},
	issn = {0035-8711},
	url = {https://ui.adsabs.harvard.edu/abs/2025MNRAS.537.2214L},
	doi = {10.1093/mnras/staf182},
	urldate = {2026-04-18},
	journal = {Monthly Notices of the Royal Astronomical Society},
	publisher = {OUP},
	author = {Li, Yuqi and Bonsor, Amy and Shorttle, Oliver and Rogers, Laura K.},
	month = feb,
	year = {2025},
	note = {ADS Bibcode: 2025MNRAS.537.2214L},
	pages = {2214--2231},
}

@article{trani_ominous_2020,
	title = {The ominous fate of exomoons around hot {Jupiters} in the high-eccentricity migration scenario},
	volume = {499},
	issn = {0035-8711},
	url = {https://ui.adsabs.harvard.edu/abs/2020MNRAS.499.4195T},
	doi = {10.1093/mnras/staa3098},
	urldate = {2026-04-18},
	journal = {Monthly Notices of the Royal Astronomical Society},
	publisher = {OUP},
	author = {Trani, Alessandro A. and Hamers, Adrian S. and Geller, Aaron and Spera, Mario},
	month = dec,
	year = {2020},
	note = {ADS Bibcode: 2020MNRAS.499.4195T},
	pages = {4195--4205},
}

@article{mathieu_blue_2025,
	title = {Blue {Stragglers} and {Friends}: {Initial} {Evolutionary} {Pathways} in {Close} {Low}-{Mass} {Binaries}},
	volume = {63},
	issn = {0066-4146},
	shorttitle = {Blue {Stragglers} and {Friends}},
	url = {https://ui.adsabs.harvard.edu/abs/2025ARA&A..63..467M},
	doi = {10.1146/annurev-astro-071221-054402},
	urldate = {2026-04-17},
	journal = {Annual Review of Astronomy and Astrophysics},
	author = {Mathieu, Robert D. and Pols, Onno R.},
	month = aug,
	year = {2025},
	note = {ADS Bibcode: 2025ARA\&A..63..467M},
	pages = {467--512},
}

@misc{esseldeurs_competition_2026,
	title = {Competition between gravity waves excited by convection and tides in stars that host a companion},
	url = {https://ui.adsabs.harvard.edu/abs/2026arXiv260326577E},
	doi = {10.48550/arXiv.2603.26577},
	urldate = {2026-04-15},
	publisher = {arXiv},
	author = {Esseldeurs, M. and Ahuir, J. and Amard, L. and Mathis, S. and Decin, L.},
	month = mar,
	year = {2026},
	note = {ADS Bibcode: 2026arXiv260326577E},
}

@article{behmard_planet_2023,
	title = {Planet engulfment signatures in twin stars},
	volume = {518},
	issn = {0035-8711},
	url = {https://ui.adsabs.harvard.edu/abs/2023MNRAS.518.5465B},
	doi = {10.1093/mnras/stac3435},
	urldate = {2026-03-04},
	journal = {Monthly Notices of the Royal Astronomical Society},
	publisher = {OUP},
	author = {Behmard, Aida and Sevilla, Jason and Fuller, Jim},
	month = feb,
	year = {2023},
	note = {ADS Bibcode: 2023MNRAS.518.5465B},
	pages = {5465--5474},
}

@article{goldstein_contour_2020,
	title = {The {Contour} {Method}: a {New} {Approach} to {Finding} {Modes} of {Nonadiabatic} {Stellar} {Pulsations}},
	volume = {899},
	issn = {0004-637X},
	shorttitle = {The {Contour} {Method}},
	url = {https://ui.adsabs.harvard.edu/abs/2020ApJ...899..116G},
	doi = {10.3847/1538-4357/aba748},
	urldate = {2026-03-04},
	journal = {The Astrophysical Journal},
	publisher = {IOP},
	author = {Goldstein, J. and Townsend, R. H. D.},
	month = aug,
	year = {2020},
	note = {ADS Bibcode: 2020ApJ...899..116G},
	pages = {116},
}

@article{townsend_gyre_2013,
	title = {{GYRE}: an open-source stellar oscillation code based on a new {Magnus} {Multiple} {Shooting} scheme},
	volume = {435},
	issn = {0035-8711},
	shorttitle = {{GYRE}},
	url = {https://ui.adsabs.harvard.edu/abs/2013MNRAS.435.3406T},
	doi = {10.1093/mnras/stt1533},
	urldate = {2026-03-04},
	journal = {Monthly Notices of the Royal Astronomical Society},
	publisher = {OUP},
	author = {Townsend, R. H. D. and Teitler, S. A.},
	month = nov,
	year = {2013},
	note = {ADS Bibcode: 2013MNRAS.435.3406T},
	pages = {3406--3418},
}

@article{jermyn_modules_2023,
	title = {Modules for {Experiments} in {Stellar} {Astrophysics} ({MESA}): {Time}-dependent {Convection}, {Energy} {Conservation}, {Automatic} {Differentiation}, and {Infrastructure}},
	volume = {265},
	issn = {0067-0049},
	shorttitle = {Modules for {Experiments} in {Stellar} {Astrophysics} ({MESA})},
	url = {https://ui.adsabs.harvard.edu/abs/2023ApJS..265...15J},
	doi = {10.3847/1538-4365/acae8d},
	urldate = {2026-03-04},
	journal = {The Astrophysical Journal Supplement Series},
	publisher = {IOP},
	author = {Jermyn, Adam S. and Bauer, Evan B. and Schwab, Josiah and Farmer, R. and Ball, Warrick H. and Bellinger, Earl P. and Dotter, Aaron and Joyce, Meridith and Marchant, Pablo and Mombarg, Joey S. G. and Wolf, William M. and Sunny Wong, Tin Long and Cinquegrana, Giulia C. and Farrell, Eoin and Smolec, R. and Thoul, Anne and Cantiello, Matteo and Herwig, Falk and Toloza, Odette and Bildsten, Lars and Townsend, Richard H. D. and Timmes, F. X.},
	month = mar,
	year = {2023},
	note = {ADS Bibcode: 2023ApJS..265...15J},
	pages = {15},
}

@article{paxton_modules_2018,
	title = {Modules for {Experiments} in {Stellar} {Astrophysics} ({MESA}): {Convective} {Boundaries}, {Element} {Diffusion}, and {Massive} {Star} {Explosions}},
	volume = {234},
	issn = {0067-0049},
	shorttitle = {Modules for {Experiments} in {Stellar} {Astrophysics} ({MESA})},
	url = {https://ui.adsabs.harvard.edu/abs/2018ApJS..234...34P},
	doi = {10.3847/1538-4365/aaa5a8},
	urldate = {2026-03-04},
	journal = {The Astrophysical Journal Supplement Series},
	publisher = {IOP},
	author = {Paxton, Bill and Schwab, Josiah and Bauer, Evan B. and Bildsten, Lars and Blinnikov, Sergei and Duffell, Paul and Farmer, R. and Goldberg, Jared A. and Marchant, Pablo and Sorokina, Elena and Thoul, Anne and Townsend, Richard H. D. and Timmes, F. X.},
	month = feb,
	year = {2018},
	note = {ADS Bibcode: 2018ApJS..234...34P},
	pages = {34},
}

@article{eggleton_aproximations_1983,
	title = {Aproximations to the radii of {Roche} lobes.},
	volume = {268},
	issn = {0004-637X},
	url = {https://ui.adsabs.harvard.edu/abs/1983ApJ...268..368E},
	doi = {10.1086/160960},
	urldate = {2026-03-04},
	journal = {The Astrophysical Journal},
	publisher = {IOP},
	author = {Eggleton, P. P.},
	month = may,
	year = {1983},
	note = {ADS Bibcode: 1983ApJ...268..368E},
	pages = {368--369},
}

@article{terquem_tidal_1998,
	title = {On the {Tidal} {Interaction} of a {Solar}-{Type} {Star} with an {Orbiting} {Companion}: {Excitation} of g-{Mode} {Oscillation} and {Orbital} {Evolution}},
	shorttitle = {On the {Tidal} {Interaction} of a {Solar}-{Type} {Star} with an {Orbiting} {Companion}},
	url = {https://iopscience.iop.org/article/10.1086/305927},
	doi = {10.1086/305927},
	language = {en},
	urldate = {2026-02-26},
	journal = {The Astrophysical Journal},
	publisher = {IOP Publishing},
	author = {Terquem, C. and Papaloizou, J. C. B. and Nelson, R. P. and Lin, D. N. C.},
	month = aug,
	year = {1998},
}

@article{vanderburg_giant_2020,
	title = {A giant planet candidate transiting a white dwarf},
	volume = {585},
	copyright = {2020 The Author(s), under exclusive licence to Springer Nature Limited},
	issn = {1476-4687},
	url = {https://www.nature.com/articles/s41586-020-2713-y},
	doi = {10.1038/s41586-020-2713-y},
	language = {en},
	number = {7825},
	urldate = {2026-03-04},
	journal = {Nature},
	publisher = {Nature Publishing Group},
	author = {Vanderburg, Andrew and Rappaport, Saul A. and Xu, Siyi and Crossfield, Ian J. M. and Becker, Juliette C. and Gary, Bruce and Murgas, Felipe and Blouin, Simon and Kaye, Thomas G. and Palle, Enric and Melis, Carl and Morris, Brett M. and Kreidberg, Laura and Gorjian, Varoujan and Morley, Caroline V. and Mann, Andrew W. and Parviainen, Hannu and Pearce, Logan A. and Newton, Elisabeth R. and Carrillo, Andreia and Zuckerman, Ben and Nelson, Lorne and Zeimann, Greg and Brown, Warren R. and Tronsgaard, René and Klein, Beth and Ricker, George R. and Vanderspek, Roland K. and Latham, David W. and Seager, Sara and Winn, Joshua N. and Jenkins, Jon M. and Adams, Fred C. and Benneke, Björn and Berardo, David and Buchhave, Lars A. and Caldwell, Douglas A. and Christiansen, Jessie L. and Collins, Karen A. and Colón, Knicole D. and Daylan, Tansu and Doty, John and Doyle, Alexandra E. and Dragomir, Diana and Dressing, Courtney and Dufour, Patrick and Fukui, Akihiko and Glidden, Ana and Guerrero, Natalia M. and Guo, Xueying and Heng, Kevin and Henriksen, Andreea I. and Huang, Chelsea X. and Kaltenegger, Lisa and Kane, Stephen R. and Lewis, John A. and Lissauer, Jack J. and Morales, Farisa and Narita, Norio and Pepper, Joshua and Rose, Mark E. and Smith, Jeffrey C. and Stassun, Keivan G. and Yu, Liang},
	month = sep,
	year = {2020},
	pages = {363--367},
}

@article{steele_nltt_2013,
	title = {{NLTT} 5306: the shortest period detached white dwarf+brown dwarf binary},
	volume = {429},
	issn = {0035-8711},
	shorttitle = {{NLTT} 5306},
	url = {https://ui.adsabs.harvard.edu/abs/2013MNRAS.429.3492S},
	doi = {10.1093/mnras/sts620},
	urldate = {2026-03-04},
	journal = {Monthly Notices of the Royal Astronomical Society},
	publisher = {OUP},
	author = {Steele, P. R. and Saglia, R. P. and Burleigh, M. R. and Marsh, T. R. and Gänsicke, B. T. and Lawrie, K. and Cappetta, M. and Girven, J. and Napiwotzki, R.},
	month = mar,
	year = {2013},
	note = {ADS Bibcode: 2013MNRAS.429.3492S},
	pages = {3492--3500},
}

@article{burleigh_near-infrared_2006,
	title = {A near-infrared spectroscopic detection of the brown dwarf in the post common envelope binary {WD} 0137-349},
	volume = {373},
	issn = {1745-3933},
	url = {https://onlinelibrary.wiley.com/doi/abs/10.1111/j.1745-3933.2006.00242.x},
	doi = {10.1111/j.1745-3933.2006.00242.x},
	language = {en},
	number = {1},
	urldate = {2026-03-04},
	journal = {Monthly Notices of the Royal Astronomical Society: Letters},
	author = {Burleigh, M. R. and Hogan, E. and Dobbie, P. D. and Napiwotzki, R. and Maxted, P. F. L.},
	year = {2006},
	note = {\_eprint: https://onlinelibrary.wiley.com/doi/pdf/10.1111/j.1745-3933.2006.00242.x},
	pages = {L55--L59},
}

@article{freedman_gaseous_2014,
	title = {{GASEOUS} {MEAN} {OPACITIES} {FOR} {GIANT} {PLANET} {AND} {ULTRACOOL} {DWARF} {ATMOSPHERES} {OVER} {A} {RANGE} {OF} {METALLICITIES} {AND} {TEMPERATURES}},
	volume = {214},
	issn = {0067-0049},
	url = {https://doi.org/10.1088/0067-0049/214/2/25},
	doi = {10.1088/0067-0049/214/2/25},
	language = {en},
	number = {2},
	urldate = {2026-03-04},
	journal = {The Astrophysical Journal Supplement Series},
	publisher = {The American Astronomical Society},
	author = {Freedman, Richard S. and Lustig-Yaeger, Jacob and Fortney, Jonathan J. and Lupu, Roxana E. and Marley, Mark S. and Lodders, Katharina},
	month = oct,
	year = {2014},
	pages = {25},
}

@article{rasio_tidal_1996,
	title = {Tidal {Decay} of {Close} {Planetary} {Orbits}},
	volume = {470},
	issn = {0004-637X},
	url = {https://ui.adsabs.harvard.edu/abs/1996ApJ...470.1187R},
	doi = {10.1086/177941},
	urldate = {2026-03-04},
	journal = {The Astrophysical Journal},
	publisher = {IOP},
	author = {Rasio, F. A. and Tout, C. A. and Lubow, S. H. and Livio, M.},
	month = oct,
	year = {1996},
	note = {ADS Bibcode: 1996ApJ...470.1187R},
	pages = {1187},
}

@inproceedings{zahn_tidal_2008,
	address = {eprint: arXiv:0807.4870},
	title = {Tidal dissipation in binary systems},
	volume = {29},
	url = {https://ui.adsabs.harvard.edu/abs/2008EAS....29...67Z},
	doi = {10.1051/eas:0829002},
	urldate = {2026-03-03},
	booktitle = {Tidal effects in stars, planets and disks},
	publisher = {EAS Publications Series},
	author = {Zahn, J.-P.},
	month = jan,
	year = {2008},
	note = {ADS Bibcode: 2008EAS....29...67Z},
	pages = {67--90},
}

@article{barker_tidal_2020,
	title = {Tidal dissipation in evolving low-mass and solar-type stars with predictions for planetary orbital decay},
	volume = {498},
	issn = {0035-8711},
	url = {https://doi.org/10.1093/mnras/staa2405},
	doi = {10.1093/mnras/staa2405},
	number = {2},
	urldate = {2026-03-03},
	journal = {Monthly Notices of the Royal Astronomical Society},
	author = {Barker, A J},
	month = sep,
	year = {2020},
	pages = {2270--2294},
}

@article{duguid_convective_2020,
	title = {Convective turbulent viscosity acting on equilibrium tidal flows: new frequency scaling of the effective viscosity},
	volume = {497},
	issn = {0035-8711},
	shorttitle = {Convective turbulent viscosity acting on equilibrium tidal flows},
	url = {https://doi.org/10.1093/mnras/staa2216},
	doi = {10.1093/mnras/staa2216},
	number = {3},
	urldate = {2026-03-03},
	journal = {Monthly Notices of the Royal Astronomical Society},
	author = {Duguid, Craig D and Barker, Adrian J and Jones, C A},
	month = sep,
	year = {2020},
	pages = {3400--3417},
}

@article{goldreich_turbulent_1977,
	title = {Turbulent {Viscosity} and {Jupiter}'s {Tidal} {Q}},
	volume = {30},
	issn = {0019-1035},
	url = {https://ui.adsabs.harvard.edu/abs/1977Icar...30..301G},
	doi = {10.1016/0019-1035(77)90163-4},
	urldate = {2026-02-28},
	journal = {Icarus},
	publisher = {Elsevier},
	author = {Goldreich, P. and Nicholson, P. D.},
	month = feb,
	year = {1977},
	note = {ADS Bibcode: 1977Icar...30..301G},
	pages = {301--304},
}

@article{zahn_marees_1966,
	title = {Les marées dans une étoile double serrée (suite)},
	volume = {29},
	issn = {0365-0499},
	url = {https://ui.adsabs.harvard.edu/abs/1966AnAp...29..489Z},
	urldate = {2026-02-28},
	journal = {Annales d'Astrophysique},
	author = {Zahn, J. P.},
	month = feb,
	year = {1966},
	note = {ADS Bibcode: 1966AnAp...29..489Z},
	pages = {489},
}

@article{hurley_evolution_2002,
	title = {Evolution of binary stars and the effect of tides on binary populations},
	volume = {329},
	issn = {0035-8711},
	url = {https://ui.adsabs.harvard.edu/abs/2002MNRAS.329..897H},
	doi = {10.1046/j.1365-8711.2002.05038.x},
	urldate = {2026-02-28},
	journal = {Monthly Notices of the Royal Astronomical Society},
	publisher = {OUP},
	author = {Hurley, Jarrod R. and Tout, Christopher A. and Pols, Onno R.},
	month = feb,
	year = {2002},
	note = {ADS Bibcode: 2002MNRAS.329..897H},
	pages = {897--928},
}

@article{witte_orbital_2002,
	title = {Orbital evolution by dynamical tides in solar type stars - {Application} to binary stars and planetary orbits {\textbar} {Astronomy} \& {Astrophysics} ({A}\&{A})},
	volume = {386},
	copyright = {© ESO, 2002},
	issn = {0004-6361},
	url = {https://www.aanda.org/articles/aa/abs/2002/16/aah3164/aah3164.html},
	language = {en-gb},
	number = {1},
	urldate = {2026-02-26},
	journal = {Astronomy \& Astrophysics},
	author = {Witte, M. G. and Savonije, G. J.},
	year = {2002},
	pages = {222--236},
}

@article{ma_orbital_2021,
	title = {Orbital {Decay} of {Short}-period {Exoplanets} via {Tidal} {Resonance} {Locking}},
	volume = {918},
	issn = {0004-637X},
	url = {https://doi.org/10.3847/1538-4357/ac088e},
	doi = {10.3847/1538-4357/ac088e},
	language = {en},
	number = {1},
	urldate = {2026-02-26},
	journal = {The Astrophysical Journal},
	publisher = {The American Astronomical Society},
	author = {Ma, Linhao and Fuller, Jim},
	month = aug,
	year = {2021},
	pages = {16},
}

@article{mirouh_detailed_2023,
	title = {Detailed equilibrium and dynamical tides: impact on circularization and synchronization in open clusters},
	volume = {524},
	issn = {0035-8711},
	shorttitle = {Detailed equilibrium and dynamical tides},
	url = {https://doi.org/10.1093/mnras/stad2048},
	doi = {10.1093/mnras/stad2048},
	number = {3},
	urldate = {2026-02-26},
	journal = {Monthly Notices of the Royal Astronomical Society},
	author = {Mirouh, Giovanni M and Hendriks, David D and Dykes, Sophie and Moe, Maxwell and Izzard, Robert G},
	month = sep,
	year = {2023},
	pages = {3978--3999},
}

@article{weinberg_nonlinear_2012,
	title = {{NONLINEAR} {TIDES} {IN} {CLOSE} {BINARY} {SYSTEMS}},
	volume = {751},
	issn = {0004-637X},
	url = {https://doi.org/10.1088/0004-637X/751/2/136},
	doi = {10.1088/0004-637X/751/2/136},
	language = {en},
	number = {2},
	urldate = {2026-02-26},
	journal = {The Astrophysical Journal},
	publisher = {The American Astronomical Society},
	author = {Weinberg, Nevin N. and Arras, Phil and Quataert, Eliot and Burkart, Josh},
	month = may,
	year = {2012},
	pages = {136},
}

@article{goodman_dynamical_1998,
	title = {Dynamical {Tide} in {Solar}-{Type} {Binaries}},
	volume = {507},
	issn = {0004-637X},
	url = {https://ui.adsabs.harvard.edu/abs/1998ApJ...507..938G},
	doi = {10.1086/306348},
	urldate = {2026-02-26},
	journal = {The Astrophysical Journal},
	publisher = {IOP},
	author = {Goodman, Jeremy and Dickson, Eric S.},
	month = nov,
	year = {1998},
	note = {ADS Bibcode: 1998ApJ...507..938G},
	pages = {938--944},
}

@article{hughes_computation_1981,
	title = {The {Computation} of {Tables} of {Hansen} {Coefficients}},
	volume = {25},
	issn = {0008-8714},
	url = {https://ui.adsabs.harvard.edu/abs/1981CeMec..25..101H},
	doi = {10.1007/BF01301812},
	urldate = {2026-02-26},
	journal = {Celestial Mechanics},
	publisher = {Springer},
	author = {Hughes, S.},
	month = sep,
	year = {1981},
	note = {ADS Bibcode: 1981CeMec..25..101H},
	pages = {101--107},
}

@article{mosser_probing_2012,
	title = {Probing the core structure and evolution of red giants using gravity-dominated mixed modes observed with {Kepler}},
	volume = {540},
	issn = {0004-6361},
	url = {https://ui.adsabs.harvard.edu/abs/2012A&A...540A.143M},
	doi = {10.1051/0004-6361/201118519},
	urldate = {2026-02-22},
	journal = {Astronomy and Astrophysics},
	publisher = {EDP},
	author = {Mosser, B. and Goupil, M. J. and Belkacem, K. and Michel, E. and Stello, D. and Marques, J. P. and Elsworth, Y. and Barban, C. and Beck, P. G. and Bedding, T. R. and De Ridder, J. and García, R. A. and Hekker, S. and Kallinger, T. and Samadi, R. and Stumpe, M. C. and Barclay, T. and Burke, C. J.},
	month = apr,
	year = {2012},
	note = {ADS Bibcode: 2012A\&A...540A.143M},
	pages = {A143},
}

@article{bedding_gravity_2011,
	title = {Gravity modes as a way to distinguish between hydrogen- and helium-burning red giant stars},
	volume = {471},
	copyright = {2011 Springer Nature Limited},
	issn = {1476-4687},
	url = {https://www.nature.com/articles/nature09935},
	doi = {10.1038/nature09935},
	language = {en},
	number = {7340},
	urldate = {2026-02-22},
	journal = {Nature},
	publisher = {Nature Publishing Group},
	author = {Bedding, Timothy R. and Mosser, Benoit and Huber, Daniel and Montalbán, Josefina and Beck, Paul and Christensen-Dalsgaard, Jørgen and Elsworth, Yvonne P. and García, Rafael A. and Miglio, Andrea and Stello, Dennis and White, Timothy R. and De Ridder, Joris and Hekker, Saskia and Aerts, Conny and Barban, Caroline and Belkacem, Kevin and Broomhall, Anne-Marie and Brown, Timothy M. and Buzasi, Derek L. and Carrier, Fabien and Chaplin, William J. and Di Mauro, Maria Pia and Dupret, Marc-Antoine and Frandsen, Søren and Gilliland, Ronald L. and Goupil, Marie-Jo and Jenkins, Jon M. and Kallinger, Thomas and Kawaler, Steven and Kjeldsen, Hans and Mathur, Savita and Noels, Arlette and Aguirre, Victor Silva and Ventura, Paolo},
	month = mar,
	year = {2011},
	pages = {608--611},
}

@article{dupret_theoretical_2009,
	title = {Theoretical amplitudes and lifetimes of non-radial solar-like oscillations in red giants},
	volume = {506},
	issn = {0004-6361},
	url = {https://ui.adsabs.harvard.edu/abs/2009A&A...506...57D},
	doi = {10.1051/0004-6361/200911713},
	urldate = {2026-02-22},
	journal = {Astronomy and Astrophysics},
	publisher = {EDP},
	author = {Dupret, M.-A. and Belkacem, K. and Samadi, R. and Montalban, J. and Moreira, O. and Miglio, A. and Godart, M. and Ventura, P. and Ludwig, H.-G. and Grigahcène, A. and Goupil, M.-J. and Noels, A. and Caffau, E.},
	month = oct,
	year = {2009},
	note = {ADS Bibcode: 2009A\&A...506...57D},
	pages = {57--67},
}

@article{narayan_wiyn_2026,
	title = {{WIYN} {Open} {Cluster} {Study}. {XCVII}. {An} {Extended} {Radial}-velocity {Survey} and {Spectroscopic} {Binary} {Orbits} in the {Open} {Cluster} {NGC} 188},
	volume = {171},
	issn = {1538-3881},
	url = {https://doi.org/10.3847/1538-3881/ae2d14},
	doi = {10.3847/1538-3881/ae2d14},
	language = {en},
	number = {2},
	urldate = {2026-01-21},
	journal = {The Astronomical Journal},
	publisher = {The American Astronomical Society},
	author = {Narayan, Ritvik Sai and Linck, Evan and Mathieu, Robert D. and Geller, Aaron M.},
	month = jan,
	year = {2026},
	pages = {102},
}

@incollection{reimers_circumstellar_1975,
	title = {Circumstellar envelopes and mass loss of red giant stars.},
	url = {https://ui.adsabs.harvard.edu/abs/1975psae.book..229R},
	urldate = {2026-01-12},
	booktitle = {Problems in stellar atmospheres and envelopes.},
	publisher = {Springer Nature},
	author = {Reimers, D.},
	month = jan,
	year = {1975},
	note = {ADS Bibcode: 1975psae.book..229R},
	pages = {229--256},
}

@article{sun_numerical_2025,
	title = {Numerical {Nonadiabatic} {Tidal} {Calculations} with {GYRE}-tides: {The} {WASP}-12 {Test} {Case}},
	volume = {995},
	issn = {0004-637X},
	shorttitle = {Numerical {Nonadiabatic} {Tidal} {Calculations} with {GYRE}-tides},
	url = {https://ui.adsabs.harvard.edu/abs/2025ApJ...995L..29S},
	doi = {10.3847/2041-8213/ae23cf},
	urldate = {2026-01-12},
	journal = {The Astrophysical Journal},
	publisher = {IOP},
	author = {Sun, Meng and Townsend, R. H. D. and Xia, Hongbo and Liu, Jifeng},
	month = dec,
	year = {2025},
	note = {ADS Bibcode: 2025ApJ...995L..29S},
	pages = {L29},
}

@article{paxton_modules_2019,
	title = {Modules for {Experiments} in {Stellar} {Astrophysics} ({MESA}): {Pulsating} {Variable} {Stars}, {Rotation}, {Convective} {Boundaries}, and {Energy} {Conservation}},
	volume = {243},
	issn = {0067-0049},
	shorttitle = {Modules for {Experiments} in {Stellar} {Astrophysics} ({MESA})},
	url = {https://ui.adsabs.harvard.edu/abs/2019ApJS..243...10P},
	doi = {10.3847/1538-4365/ab2241},
	urldate = {2026-01-12},
	journal = {The Astrophysical Journal Supplement Series},
	publisher = {IOP},
	author = {Paxton, Bill and Smolec, R. and Schwab, Josiah and Gautschy, A. and Bildsten, Lars and Cantiello, Matteo and Dotter, Aaron and Farmer, R. and Goldberg, Jared A. and Jermyn, Adam S. and Kanbur, S. M. and Marchant, Pablo and Thoul, Anne and Townsend, Richard H. D. and Wolf, William M. and Zhang, Michael and Timmes, F. X.},
	month = jul,
	year = {2019},
	note = {ADS Bibcode: 2019ApJS..243...10P},
	pages = {10},
}

@article{paxton_modules_2015,
	title = {Modules for {Experiments} in {Stellar} {Astrophysics} ({MESA}): {Binaries}, {Pulsations}, and {Explosions}},
	volume = {220},
	issn = {0067-0049},
	shorttitle = {Modules for {Experiments} in {Stellar} {Astrophysics} ({MESA})},
	url = {https://ui.adsabs.harvard.edu/abs/2015ApJS..220...15P},
	doi = {10.1088/0067-0049/220/1/15},
	urldate = {2025-04-06},
	journal = {The Astrophysical Journal Supplement Series},
	publisher = {IOP},
	author = {Paxton, Bill and Marchant, Pablo and Schwab, Josiah and Bauer, Evan B. and Bildsten, Lars and Cantiello, Matteo and Dessart, Luc and Farmer, R. and Hu, H. and Langer, N. and Townsend, R. H. D. and Townsley, Dean M. and Timmes, F. X.},
	month = sep,
	year = {2015},
	note = {ADS Bibcode: 2015ApJS..220...15P},
	pages = {15},
}

@article{paxton_modules_2011,
	title = {Modules for {Experiments} in {Stellar} {Astrophysics} ({MESA})},
	volume = {192},
	issn = {0067-0049},
	url = {https://ui.adsabs.harvard.edu/abs/2011ApJS..192....3P},
	doi = {10.1088/0067-0049/192/1/3},
	urldate = {2025-04-06},
	journal = {The Astrophysical Journal Supplement Series},
	publisher = {IOP},
	author = {Paxton, Bill and Bildsten, Lars and Dotter, Aaron and Herwig, Falk and Lesaffre, Pierre and Timmes, Frank},
	month = jan,
	year = {2011},
	note = {ADS Bibcode: 2011ApJS..192....3P},
	pages = {3},
}

@article{paxton_modules_2013,
	title = {Modules for {Experiments} in {Stellar} {Astrophysics} ({MESA}): {Planets}, {Oscillations}, {Rotation}, and {Massive} {Stars}},
	volume = {208},
	issn = {0067-0049},
	shorttitle = {Modules for {Experiments} in {Stellar} {Astrophysics} ({MESA})},
	url = {https://ui.adsabs.harvard.edu/abs/2013ApJS..208....4P},
	doi = {10.1088/0067-0049/208/1/4},
	urldate = {2025-04-06},
	journal = {The Astrophysical Journal Supplement Series},
	publisher = {IOP},
	author = {Paxton, Bill and Cantiello, Matteo and Arras, Phil and Bildsten, Lars and Brown, Edward F. and Dotter, Aaron and Mankovich, Christopher and Montgomery, M. H. and Stello, Dennis and Timmes, F. X. and Townsend, Richard},
	month = sep,
	year = {2013},
	note = {ADS Bibcode: 2013ApJS..208....4P},
	pages = {4},
}

@article{ogilvie_tidal_2014,
	title = {Tidal {Dissipation} in {Stars} and {Giant} {Planets}},
	volume = {52},
	issn = {0066-4146},
	url = {https://ui.adsabs.harvard.edu/abs/2014ARA&A..52..171O},
	doi = {10.1146/annurev-astro-081913-035941},
	urldate = {2025-12-30},
	journal = {Annual Review of Astronomy and Astrophysics},
	author = {Ogilvie, Gordon I.},
	month = aug,
	year = {2014},
	note = {ADS Bibcode: 2014ARA\&A..52..171O},
	pages = {171--210},
}

@article{willems_energy_2010,
	title = {Energy {Dissipation} {Through} {Quasi}-static {Tides} in {White} {Dwarf} {Binaries}},
	volume = {713},
	issn = {0004-637X},
	url = {https://ui.adsabs.harvard.edu/abs/2010ApJ...713..239W},
	doi = {10.1088/0004-637X/713/1/239},
	urldate = {2025-12-27},
	journal = {The Astrophysical Journal},
	publisher = {IOP},
	author = {Willems, B. and Deloye, C. J. and Kalogera, V.},
	month = apr,
	year = {2010},
	note = {ADS Bibcode: 2010ApJ...713..239W},
	pages = {239--256},
}

@article{townsend_angular_2018,
	title = {Angular momentum transport by heat-driven g-modes in slowly pulsating {B} stars},
	volume = {475},
	issn = {0035-8711},
	url = {https://ui.adsabs.harvard.edu/abs/2018MNRAS.475..879T},
	doi = {10.1093/mnras/stx3142},
	urldate = {2025-12-27},
	journal = {Monthly Notices of the Royal Astronomical Society},
	publisher = {OUP},
	author = {Townsend, R. H. D. and Goldstein, J. and Zweibel, E. G.},
	month = mar,
	year = {2018},
	note = {ADS Bibcode: 2018MNRAS.475..879T},
	pages = {879--893},
}

@article{nordhaus_tides_2010,
	title = {Tides and tidal engulfment in post-main-sequence binaries: period gaps for planets and brown dwarfs around white dwarfs},
	volume = {408},
	issn = {0035-8711},
	shorttitle = {Tides and tidal engulfment in post-main-sequence binaries},
	url = {https://ui.adsabs.harvard.edu/abs/2010MNRAS.408..631N},
	doi = {10.1111/j.1365-2966.2010.17155.x},
	urldate = {2025-12-08},
	journal = {Monthly Notices of the Royal Astronomical Society},
	publisher = {OUP},
	author = {Nordhaus, J. and Spiegel, D. S. and Ibgui, L. and Goodman, J. and Burrows, A.},
	month = oct,
	year = {2010},
	note = {ADS Bibcode: 2010MNRAS.408..631N},
	pages = {631--641},
}

@article{sun_gyre_tides_2023,
	title = {gyre\_tides: {Modeling} {Binary} {Tides} within the {GYRE} {Stellar} {Oscillation} {Code}},
	volume = {945},
	issn = {0004-637X},
	shorttitle = {gyre\_tides},
	url = {https://ui.adsabs.harvard.edu/abs/2023ApJ...945...43S},
	doi = {10.3847/1538-4357/acb33a},
	urldate = {2025-12-08},
	journal = {The Astrophysical Journal},
	publisher = {IOP},
	author = {Sun, Meng and Townsend, R. H. D. and Guo, Zhao},
	month = mar,
	year = {2023},
	note = {ADS Bibcode: 2023ApJ...945...43S},
	pages = {43},
}

@article{barker_internal_2010,
	title = {On internal wave breaking and tidal dissipation near the centre of a solar-type star},
	volume = {404},
	issn = {0035-8711},
	url = {https://ui.adsabs.harvard.edu/abs/2010MNRAS.404.1849B},
	doi = {10.1111/j.1365-2966.2010.16400.x},
	urldate = {2025-12-07},
	journal = {Monthly Notices of the Royal Astronomical Society},
	publisher = {OUP},
	author = {Barker, Adrian J. and Ogilvie, Gordon I.},
	month = jun,
	year = {2010},
	note = {ADS Bibcode: 2010MNRAS.404.1849B},
	pages = {1849--1868},
}

@article{ogilvie_tidal_2007,
	title = {Tidal {Dissipation} in {Rotating} {Solar}-{Type} {Stars}},
	volume = {661},
	issn = {0004-637X},
	url = {https://ui.adsabs.harvard.edu/abs/2007ApJ...661.1180O},
	doi = {10.1086/515435},
	urldate = {2025-12-07},
	journal = {The Astrophysical Journal},
	publisher = {IOP},
	author = {Ogilvie, G. I. and Lin, D. N. C.},
	month = jun,
	year = {2007},
	note = {ADS Bibcode: 2007ApJ...661.1180O},
	pages = {1180--1191},
}

@article{hut_tidal_1981,
	title = {Tidal evolution in close binary systems.},
	volume = {99},
	issn = {0004-6361},
	url = {https://ui.adsabs.harvard.edu/abs/1981A&A....99..126H},
	urldate = {2025-12-07},
	journal = {Astronomy and Astrophysics},
	publisher = {EDP},
	author = {Hut, P.},
	month = jun,
	year = {1981},
	note = {ADS Bibcode: 1981A\&A....99..126H},
	pages = {126--140},
}

@article{zahn_dynamical_1975,
	title = {The dynamical tide in close binaries.},
	volume = {41},
	issn = {0004-6361},
	url = {https://ui.adsabs.harvard.edu/abs/1975A&A....41..329Z},
	urldate = {2025-12-07},
	journal = {Astronomy and Astrophysics},
	publisher = {EDP},
	author = {Zahn, J.-P.},
	month = jul,
	year = {1975},
	note = {ADS Bibcode: 1975A\&A....41..329Z},
	pages = {329--344},
}

@article{zahn_tidal_1989,
	title = {Tidal evolution of close binary stars. {I} - {Revisiting} the theory of the equilibrium tide},
	volume = {220},
	issn = {0004-6361},
	url = {https://ui.adsabs.harvard.edu/abs/1989A&A...220..112Z},
	urldate = {2025-12-07},
	journal = {Astronomy and Astrophysics},
	publisher = {EDP},
	author = {Zahn, J.-P.},
	month = aug,
	year = {1989},
	note = {ADS Bibcode: 1989A\&A...220..112Z},
	pages = {112--116},
}

@article{oconnor_giant_2023,
	title = {Giant {Planet} {Engulfment} by {Evolved} {Giant} {Stars}: {Light} {Curves}, {Asteroseismology}, and {Survivability}},
	volume = {950},
	issn = {0004-637X},
	shorttitle = {Giant {Planet} {Engulfment} by {Evolved} {Giant} {Stars}},
	url = {https://ui.adsabs.harvard.edu/abs/2023ApJ...950..128O},
	doi = {10.3847/1538-4357/acd2d4},
	urldate = {2025-12-07},
	journal = {The Astrophysical Journal},
	publisher = {IOP},
	author = {O'Connor, Christopher E. and Bildsten, Lars and Cantiello, Matteo and Lai, Dong},
	month = jun,
	year = {2023},
	note = {ADS Bibcode: 2023ApJ...950..128O},
	pages = {128},
}

@article{yarza_hydrodynamics_2023,
	title = {Hydrodynamics and {Survivability} during {Post}-main-sequence {Planetary} {Engulfment}},
	volume = {954},
	issn = {0004-637X},
	url = {https://ui.adsabs.harvard.edu/abs/2023ApJ...954..176Y},
	doi = {10.3847/1538-4357/acbdfc},
	urldate = {2025-12-07},
	journal = {The Astrophysical Journal},
	publisher = {IOP},
	author = {Yarza, Ricardo and Razo-López, Naela B. and Murguia-Berthier, Ariadna and Everson, Rosa Wallace and Antoni, Andrea and MacLeod, Morgan and Soares-Furtado, Melinda and Lee, Dongwook and Ramirez-Ruiz, Enrico},
	month = sep,
	year = {2023},
	note = {ADS Bibcode: 2023ApJ...954..176Y},
	pages = {176},
}

@article{soares-furtado_lithium_2021,
	title = {Lithium {Enrichment} {Signatures} of {Planetary} {Engulfment} {Events} in {Evolved} {Stars}},
	volume = {162},
	issn = {0004-6256},
	url = {https://ui.adsabs.harvard.edu/abs/2021AJ....162..273S},
	doi = {10.3847/1538-3881/ac273c},
	urldate = {2025-12-07},
	journal = {The Astronomical Journal},
	publisher = {IOP},
	author = {Soares-Furtado, Melinda and Cantiello, Matteo and MacLeod, Morgan and Ness, Melissa K.},
	month = dec,
	year = {2021},
	note = {ADS Bibcode: 2021AJ....162..273S},
	pages = {273},
}

@article{oetjens_influence_2020,
	title = {The influence of planetary engulfment on stellar rotation in metal-poor main-sequence stars},
	volume = {643},
	issn = {0004-6361},
	url = {https://ui.adsabs.harvard.edu/abs/2020A&A...643A..34O},
	doi = {10.1051/0004-6361/202038653},
	urldate = {2025-12-07},
	journal = {Astronomy and Astrophysics},
	publisher = {EDP},
	author = {Oetjens, A. and Carone, L. and Bergemann, M. and Serenelli, A.},
	month = nov,
	year = {2020},
	note = {ADS Bibcode: 2020A\&A...643A..34O},
	pages = {A34},
}

@article{tayar_spinning_2022,
	title = {Spinning up the {Surface}: {Evidence} for {Planetary} {Engulfment} or {Unexpected} {Angular} {Momentum} {Transport}?},
	volume = {940},
	issn = {0004-637X},
	shorttitle = {Spinning up the {Surface}},
	url = {https://ui.adsabs.harvard.edu/abs/2022ApJ...940...23T},
	doi = {10.3847/1538-4357/ac9312},
	urldate = {2025-12-07},
	journal = {The Astrophysical Journal},
	publisher = {IOP},
	author = {Tayar, Jamie and Moyano, Facundo D. and Soares-Furtado, Melinda and Escorza, Ana and Joyce, Meridith and Martell, Sarah L. and García, Rafael A. and Breton, Sylvain N. and Mathis, Stéphane and Mathur, Savita and Delsanti, Vincent and Kiefer, Sven and Reffert, Sabine and Bowman, Dominic M. and Van Reeth, Timothy and Shetye, Shreeya and Gehan, Charlotte and Grunblatt, Samuel K.},
	month = nov,
	year = {2022},
	note = {ADS Bibcode: 2022ApJ...940...23T},
	pages = {23},
}

@inproceedings{danchi_evolution_2006,
	title = {Evolution of the {Habitable} {Zone} and {Search} for {Life} {Around} {Giant} {Stars} {Part} {II}: {Feasibility} with {Darwin}/{TPF}},
	shorttitle = {Evolution of the {Habitable} {Zone} and {Search} for {Life} {Around} {Giant} {Stars} {Part} {II}},
	url = {https://ui.adsabs.harvard.edu/abs/2006dies.conf...65D},
	doi = {10.1017/S1743921306009094},
	urldate = {2025-12-07},
	booktitle = {{AU} {Colloquium} 200},
	author = {Danchi, W. C. and Lopez, B. and Schneider, J. and Belu, A. and Barry, R. and Rajagopal, J. and Richardson, L. J.},
	month = jan,
	year = {2006},
	note = {ADS Bibcode: 2006dies.conf...65D},
	pages = {65--70},
}

@article{danchi_effect_2013,
	title = {Effect of {Metallicity} on the {Evolution} of the {Habitable} {Zone} from the {Pre}-main {Sequence} to the {Asymptotic} {Giant} {Branch} and the {Search} for {Life}},
	volume = {769},
	issn = {0004-637X},
	url = {https://ui.adsabs.harvard.edu/abs/2013ApJ...769...27D},
	doi = {10.1088/0004-637X/769/1/27},
	urldate = {2025-12-07},
	journal = {The Astrophysical Journal},
	publisher = {IOP},
	author = {Danchi, William C. and Lopez, Bruno},
	month = may,
	year = {2013},
	note = {ADS Bibcode: 2013ApJ...769...27D},
	pages = {27},
}

@article{ramirez_habitable_2016,
	title = {Habitable {Zones} of {Post}-{Main} {Sequence} {Stars}},
	volume = {823},
	issn = {0004-637X},
	url = {https://ui.adsabs.harvard.edu/abs/2016ApJ...823....6R},
	doi = {10.3847/0004-637X/823/1/6},
	urldate = {2025-12-07},
	journal = {The Astrophysical Journal},
	publisher = {IOP},
	author = {Ramirez, Ramses M. and Kaltenegger, Lisa},
	month = may,
	year = {2016},
	note = {ADS Bibcode: 2016ApJ...823....6R},
	pages = {6},
}

@article{veras_post-main-sequence_2016,
	title = {Post-main-sequence planetary system evolution},
	volume = {3},
	url = {https://ui.adsabs.harvard.edu/abs/2016RSOS....350571V},
	doi = {10.1098/rsos.150571},
	urldate = {2025-12-07},
	journal = {Royal Society Open Science},
	author = {Veras, Dimitri},
	month = feb,
	year = {2016},
	note = {ADS Bibcode: 2016RSOS....350571V},
	pages = {150571},
}

@article{villaver_orbital_2009,
	title = {The {Orbital} {Evolution} of {Gas} {Giant} {Planets} {Around} {Giant} {Stars}},
	volume = {705},
	issn = {0004-637X},
	url = {https://ui.adsabs.harvard.edu/abs/2009ApJ...705L..81V},
	doi = {10.1088/0004-637X/705/1/L81},
	urldate = {2025-12-07},
	journal = {The Astrophysical Journal},
	publisher = {IOP},
	author = {Villaver, Eva and Livio, Mario},
	month = nov,
	year = {2009},
	note = {ADS Bibcode: 2009ApJ...705L..81V},
	pages = {L81--L85},
}

@article{penev_direct_2009,
	title = {Direct {Calculation} of the {Turbulent} {Dissipation} {Efficiency} in {Anelastic} {Convection}},
	volume = {705},
	issn = {0004-637X},
	url = {https://ui.adsabs.harvard.edu/abs/2009ApJ...705..285P},
	doi = {10.1088/0004-637X/705/1/285},
	urldate = {2025-12-07},
	journal = {The Astrophysical Journal},
	publisher = {IOP},
	author = {Penev, Kaloyan and Barranco, Joseph and Sasselov, Dimitar},
	month = nov,
	year = {2009},
	note = {ADS Bibcode: 2009ApJ...705..285P},
	pages = {285--297},
}

@article{verbunt_tidal_1995,
	title = {Tidal circularization and the eccentricity of binaries containing giant stars.},
	volume = {296},
	issn = {0004-6361},
	url = {https://ui.adsabs.harvard.edu/abs/1995A&A...296..709V},
	urldate = {2025-12-07},
	journal = {Astronomy and Astrophysics},
	publisher = {EDP},
	author = {Verbunt, F. and Phinney, E. S.},
	month = apr,
	year = {1995},
	note = {ADS Bibcode: 1995A\&A...296..709V},
	pages = {709},
}

@article{zahn_tidal_1977,
	title = {Tidal friction in close binary systems.},
	volume = {57},
	issn = {0004-6361},
	url = {https://ui.adsabs.harvard.edu/abs/1977A&A....57..383Z},
	urldate = {2025-12-07},
	journal = {Astronomy and Astrophysics},
	publisher = {EDP},
	author = {Zahn, J.-P.},
	month = may,
	year = {1977},
	note = {ADS Bibcode: 1977A\&A....57..383Z},
	pages = {383--394},
}
\bibliographystyle{aasjournalv7}

\end{CJK*}
\end{document}